\documentclass[unpublished,a4paper]{quantumarticle}
\pdfoutput=1

\usepackage{graphicx}
\usepackage{amsthm}
\usepackage{thmtools}
\usepackage{mathtools}
\usepackage{thm-restate}
\usepackage{amsmath}
\usepackage{algorithm}
\usepackage{algorithmic}
\usepackage{amssymb}
\usepackage{bm}
\usepackage{xcolor}
\usepackage{placeins}
\usepackage[normalem]{ulem}
\usepackage[colorlinks]{hyperref}
\hypersetup{
	pdfstartview={FitH},
	pdfnewwindow=true,
	colorlinks=true,
	linkcolor=blue,
	citecolor=blue,
	filecolor=blue,
	urlcolor=blue}
\usepackage{tikz}
\usetikzlibrary{calc,backgrounds}
\usepackage{physics}
\usepackage{cancel}
\usepackage{multirow}
\usepackage{pgffor}
\usepackage{url}
\usepackage{hypcap}
\usepackage{dsfont}
\usepackage{soul}

\usepackage[caption=false]{subfig}

\usepackage{hyperref}
\hypersetup{pdfpagemode=UseNone}

\allowdisplaybreaks

\newcommand{\thm}[1]{\hyperref[thm:#1]{Theorem~\ref*{thm:#1}}}
\newcommand{\defn}[1]{\hyperref[defn:#1]{Definition~\ref*{defn:#1}}}
\newcommand{\lem}[1]{\hyperref[lem:#1]{Lemma~\ref*{lem:#1}}}
\newcommand{\prop}[1]{\hyperref[prop:#1]{Proposition~\ref*{prop:#1}}}
\newcommand{\fig}[1]{\hyperref[fig:#1]{Figure~\ref*{fig:#1}}}
\newcommand{\tab}[1]{\hyperref[tab:#1]{Table~\ref*{tab:#1}}}
\renewcommand{\sec}[1]{\hyperref[sec:#1]{Section~\ref*{sec:#1}}}
\newcommand{\app}[1]{\hyperref[app:#1]{Appendix~\ref*{app:#1}}}
\newcommand{\cor}[1]{\hyperref[cor:#1]{Corollary~\ref*{cor:#1}}}
\newcommand{\obs}[1]{\hyperref[obs:#1]{Observation~\ref*{obs:#1}}}
\newcommand{\nn}{\nonumber \\}
\newcommand{\append}[1]{\hyperref[append:#1]{Appendix~\ref*{append:#1}}}

\newcommand{\WT}{W}

\usepackage{qcircuit}

\usepackage[capitalise]{cleveref} 

\newcommand{\subfigimg}[3][,]{%
  \setbox1=\hbox{\includegraphics[#1]{#3}}% Store image in box
  \leavevmode\rlap{\usebox1}% Print image
  \rlap{\hspace*{0.8\linewidth}\raisebox{\dimexpr\ht1-2\baselineskip}{#2}}% Print label
  \phantom{\usebox1}% Insert appropriate spacing
}

\newcommand{\MQ}{\affiliation{%
School of Mathematical and Physical Sciences,
Macquarie University, Sydney, NSW 2109, AU} }

\newcommand{\CQ}{\affiliation{%
ContinoQuantum, Sydney, NSW 2093, AU}}

\newcommand{\Google}{\affiliation{%
Google Quantum AI, Venice, CA 90291, USA}}

\newcommand{\UMD}{\affiliation{%
Joint Center for Quantum Information and Computer Science, University of Maryland, College Park, MD 20742, USA}}          

\begin{document}
\title{The discrete adiabatic quantum linear system solver has lower constant factors  than the randomised adiabatic solver}

\author{Pedro C. S. Costa}\email{pcs.costa@protonmail.com} \CQ\MQ
\author{Dong An} \UMD
\author{Ryan Babbush}\Google
\author{Dominic W. Berry} \MQ

\begin{abstract}
The solution of linear systems of equations is the basis of many other quantum algorithms, and recent results provided an algorithm with optimal scaling in both the condition number $\kappa$ and the allowable error $\epsilon$ [PRX Quantum \textbf{3}, 040303 (2022)].
That work was based on the discrete adiabatic theorem, and worked out an explicit constant factor for an upper bound on the complexity.
Here we show via numerical testing on random matrices that the constant factor is in practice about 1,200 times smaller than the upper bound found numerically in the previous results. That means that this approach is far more efficient than might naively be expected from the upper bound.
In particular, it is about an order of magnitude more efficient than using a randomised approach from [arXiv:2305.11352] that claimed to be more efficient.
\end{abstract}

\maketitle

\section{Introduction}

Finding solutions to a system of linear equations is of great importance in many areas of science and technology, and is the foundation for many quantum algorithms.
When we consider classical approaches for solving linear system problems, we have in the best cases a total number of steps proportional to the number of equations, which can make the solution prohibitively costly when we have a large linear systems problem.
On the other hand, quantum computers allow us to deal with the linear systems problems in these prohibitive regions by providing the output as a quantum state. The first proposal for the quantum linear system problem (QLSP) was given by Harrow, Hassidim, and Lloyd (HHL) \cite{Harrow_2009}, which was subsequently refined in later works \cite{an2022quantum,ambainis2010variable,Lin2020optimalpolynomial,CKS} with the optimal approach provided by Costa et al.\ in \cite{CostaAnYuvalEtAl2022}.

The goal of the QLSP is to efficiently produce a quantum state $\ket{x}$ proportional to the solution of $Ax=b$, where $A\in \mathbb{C}^{N\times N}$. For simplicity we assume $N=2^n$ and $\|A\|=1$, where $\|\cdot\|$ denotes the spectral norm. The ground-breaking algorithm from HHL produces the state $\ket{x}$ with cost $\order{\text{poly}(n)\kappa^2/\epsilon}$, where $\kappa=\|A\|\|A^{-1}\|$ is the condition number of $A$, and $\epsilon$ is the maximum allowable error.
Despite the exponential improvement given by HHL over the classical results on the dimension of the problem, e.g., $\order{N \sqrt{\kappa}\log(1/\epsilon)}$ achieved by the conjugate gradient descent method, there was a plenty of scope to improve on the other parameters in the complexity.

Quantum algorithms aim to provide complexity polynomial in $n$, an exponential speedup over the classical methods.
It is known \cite{RobinAram} that the lower bound for scaling of the complexity of quantum algorithms in $\kappa$ and $\epsilon$ is $\Omega(\kappa\log(1/\epsilon))$ (see \cref{app:lower} for details). This bound is in terms of calls to a block encoding for $A$, and the gate complexity will have a factor
corresponding to the cost of block encoding $A$, usually taken to be $\text{poly}(n)$ for efficient block encodings.
That shows that the optimal scaling is linear in $\kappa$ when aiming for the exponential quantum speedup in $n$, as opposed to the $\sqrt\kappa$ classical complexity.
This lower bound on the quantum complexity was achieved in \cite{CostaAnYuvalEtAl2022}, thereby providing the asymptotically optimal quantum algorithm.

The most significant improvements for the QLSP after \cite{Harrow_2009} are based on adiabatic quantum computing (AQC) \cite{an2022quantum,Lin2020,PhysRevLett.122.060504}, as opposed to the quantum phase estimation and amplitude amplification approach of HHL.
Reference \cite{CostaAnYuvalEtAl2022} is based upon AQC, but the main new feature is a discrete formulation of the adiabatic theorem, where the time evolution is based on the quantum walk operator $W_T(s)$; that enables the optimal complexity for the QLSP.
We refer to that as the quantum walk (QW) method.
In addition to the asymptotically scaling results for the QLSP in \cite{CostaAnYuvalEtAl2022}, the authors also provided strict upper bounds, which can be used to estimate the total number of quantum walk steps for a given solution error $\epsilon$.

The upper bound derived for the discrete adiabatic theorem (DAT) given in \cite{CostaAnYuvalEtAl2022}  results from a combination of several approximations from different operators. The approximations allow them to show that the total number of walk steps grows linearly in $\kappa$, i.e., $T=\alpha \kappa /\Delta$ for allowable error $\Delta$ when the parameters of the QLSP are plugged in the DAT. However, due to the loose upper bound from DAT, the value for $\alpha$ is very large.
The constant factor $\alpha= 2305$, given on page 16 of \cite{CostaAnYuvalEtAl2022}, can be used only for a rough estimate of the total number of quantum walk steps needed to yield a desired error.

An earlier proposal for solving the QLSP is from Ref.~\cite{PhysRevLett.122.060504}, which uses time evolution for randomised times in order to emulate measurements for the quantum Zeno effect.
Here, we refer to that as the Randomized method (RM).
The complexity of that approach scales as $\order{\kappa\log(\kappa/\epsilon)}$, so has an extra factor of $\log\kappa$ in the complexity over the QW method.
A later analysis in Ref.~\cite{jennings2023efficient} showed a relatively tight upper bound for the RM, and argued that it would be faster in practice than the QW method due to the smaller constant factor.

Here, we test the actual performance of the 
QW method and RM
by performing numerical simulation of the solution with many samples of randomly generated matrices. Our numerical tests show that in practice, the constant factor for the QW method is $\alpha=1.84$, which is about 1,200 times smaller than what is given in \cite{CostaAnYuvalEtAl2022}.
Moreover, our tests demonstrate that the quantum walk method is, on average, approximately 7 times more efficient than the RM. This results in an overall improvement of at least a factor of 3 in total complexity, including the filtering step for both methods, for realistic target errors $\varepsilon$ of the solution even without accounting for the additional overhead associated with simulating time evolution.

The background of the discrete quantum walk and Randomized method are presented in \cref{sec:back}.
Details about the numerical tests for the quantum walk and Randomized method are given in \cref{sec:Ne}.
In \cref{sec:filter}, we discuss the cost of implementation of the filtering method, which is applied after the walk steps, to estimate the appropriate target precision for the adiabatic evolution.
We then conclude in \cref{sec:conc}.
In appendices we provide technical details about the lower bound on the complexity, circuit construction of the walk operators, and further numerical tests.

\section{Background}
\label{sec:back}
In this section we summarise the background of how the discrete quantum walk method and Randomized method are performed.
Readers who are familiar with the background may wish to skip to the results in Section~\ref{sec:Ne}.

\subsection{Quantum walk method}
\label{sec:QW_meth}

We start this section by introducing the basic theory of adiabatic quantum computation which is independent of the application.
In AQC, we have a time-dependent system constructed from a linear interpolation between two time-independent Hamiltonians as follows
\begin{equation}
\label{eq:Ham_prob}
    H(s) = (1-f(s))H_0 + f(s)H_1,
\end{equation}
where the function $f(s): [0,1] \rightarrow [0,1]$ is called the schedule function. Normally we have an eigenpath between these Hamiltonians where the starting eigenstate of $H_0$ is easy to prepare, and the final eigenstate given by $H_1$ is the one that encodes the solution to the problem that we are trying to determine.

For linear systems solvers, the final eigenstate given by $H(1)$ should encode the normalised solution for a linear system. In other words, for $A\in \mathbb{C}^{N\times N}$ an invertible matrix with $\|A\|=1$, and a normalised vector $\ket{b}\in \mathbb{C}^N$ the goal is to prepare a normalised state $\ket{\tilde{x}}$, which is an approximation of $\ket{x}=A^{-1}\ket{b}/\|A^{-1}\ket{b}\|$.
For precision $\epsilon$ of the approximation, we require $\|\ket{\tilde{x}} - \ket{x}\|\leq \epsilon$.

In the following subsections, we show how we should encode both $H_0$ and $H_1$ when our goal is to use them for solving QLSP when $A$ is a Hermitian and positive-definite matrix, and also the case where $A$ does not have to be a Hermitian matrix.
In both cases, we also look at a family of scheduling functions for the problem that slows down the evolution according to the gap from the interpolated Hamiltonian given by \cref{eq:Ham_prob}. 

After we set up the time-dependent Hamiltonian in both scenarios of $A$, we apply the method from \cite{CostaAnYuvalEtAl2022} to transform the Hamiltonian into a discrete quantum walk, which is used as the evolution operator for the discrete adiabatic evolution.
This discrete quantum walk is to obtain an approximate solution of the linear system, then a filter is used to ensure that the solution is obtained to high accuracy.

\subsubsection{Positive definite and Hermitian matrix A}
\label{sec:PD}

As explained in \cite{CostaAnYuvalEtAl2022} we have $\ket{\tilde{x}}$ as the state achieved from the steps of the walk, and $\ket{x}$ would be obtained from the ideal adiabatic evolution. Starting with the simplest case, where $A$ is Hermitian and positive definite, one takes the Hamiltonians \cite{an2022quantum}
\begin{equation}
\label{eq:H0}
    H_0\coloneqq\begin{pmatrix} 0 &  Q_b\\
 Q_b & 0
\end{pmatrix},
\end{equation}
and
\begin{equation}
\label{eq:H1}
    H_1\coloneqq \begin{pmatrix} 0 &  AQ_b\\
 Q_bA & 0
\end{pmatrix},
\end{equation}
where $Q_b=I_N-\ket{b}\bra{b}$. The state $\ket{0,b}$ is an eigenstate of $H_0$ with eigenvalue 0, and one would aim for this to evolve adiabatically to eigenstate $\ket{0,A^{-1}b}$ of $H_1$.

Denoting the condition number of the matrix as $\kappa$, a lower bound for the gap of $H(s)$ is \cite{an2022quantum}
\begin{equation}
\label{eq:gapSc}
    \Delta_0(s)= 1- f(s) + f(s)/\kappa.
\end{equation}
As explained in \cite{an2022quantum,CostaAnYuvalEtAl2022} the goal is to provide a schedule function that slows down the evolution as the gap becomes small, according to the condition given in \cite{jansen2007bounds}
\begin{equation}
\label{eq:gapCon}
    \dot{f}(s) = d_p\Delta_0^p(s),
\end{equation}
where $f(0)=0$, $p>0$ and $d_p=\int_0^1\Delta_0^{-p}(u)\, du$ is a normalisation constant chosen so that $f(1)=1$. According to \cite{an2022quantum} the following schedule function
\begin{equation}
\label{eq:sched1}
f(s) = \frac{\kappa}{\kappa - 1}\left[1-\left(1+s\left(\kappa^{p-1}-1\right)\right)^{\frac{1}{1-p}}\right], 
\end{equation}
satisfies \cref{eq:gapCon}. %, but with $\Delta_0(s)$ replaced with the lower bound on the gap from \cref{eq:gapSc}.

Distinct from the continuous version of the adiabatic theorem, in the discrete version of the theorem, we have to take into account the smallest gap between the different groups of eigenvalues of $\WT (s)$ for $s,s+1/T$, and $s+2/T$, which are given by
\begin{equation}
\label{eq:Gaps}
    \Delta_k(s) = 1- f(s+k/T) + f(s+k/T)/\kappa, \quad k=0,1,2.
\end{equation}
The methods for block encoding the walk operator are explained in Appendix \ref{app:block}.

\subsubsection{Non-Hermitian matrix A}\label{sec:non_herm}

To address the case where $A$ is not positive definite or Hermitian, we pay an extra space cost by doubling its dimension to transform it into a Hermitian matrix before solving it in the quantum algorithm, i.e.,
\begin{equation}
\label{eq:Avec}
\mathbf{A}\coloneqq\begin{pmatrix} 0 &  A\\
 A^{\dagger} & 0
\end{pmatrix}.
\end{equation}

As described in Ref.~\cite{an2022quantum} the Hamiltonian is then constructed as
\begin{equation}
\label{eq:Hsencoding}
H(s) = \begin{pmatrix}
    0 & A(f(s)) Q_{\mathbf{b}} \\
Q_{\mathbf{b}} A(f(s)) & 0
    \end{pmatrix},
\end{equation}
where
\begin{equation}\label{eq:Af}
A(f) \coloneqq (1-f) \sigma_z \otimes I_N + f \mathbf{A} =
\begin{pmatrix} (1-f)I &  f A\\
 f A^{\dagger} & -(1-f)I
\end{pmatrix},
\end{equation}
and $Q_{\mathbf{b}}= I_{2N}-\ket{0,b}\bra{0,b}$ as the projection for this Hamiltonian formulation. As discussed in \cite{CostaAnYuvalEtAl2022} it is equivalent to taking $H(s) = (1-f(s)) H_0 + f(s) H_1$
with
\begin{align}
\label{eq:H0H1_general}
H_0&=\sigma_+ \otimes \left[(\sigma_z \otimes I_N) Q_{\mathbf{b}}\right] + \sigma_- \otimes \left[Q_{\mathbf{b}} (\sigma_z \otimes I_N)\right]\nonumber \\
    H_1&=\sigma_+ \otimes \left[\mathbf{A} Q_{\mathbf{b}}\right] + \sigma_- \otimes \left[Q_{\mathbf{b}} \mathbf{A}\right],
\end{align}
where $\sigma_+=\ket{1}\bra{0}$, $\sigma_-=\ket{0}\bra{1}$.

As in the positive-definite case, the schedule function can be determined from gap via the differential equation \eqref{eq:gapCon}.
In this case the gap is lower bounded as $\sqrt{(1-f(s))^2+(f(s)/\kappa)^2}$ so there is not a closed-form solution of \cref{eq:gapCon}.

Instead, one can use the relation that for $0\leq f(s) \leq 1$, 
\begin{equation}
\sqrt{(1-f(s))^2 + (f(s)/\kappa)^2} \geq (1-f(s) + f(s)/\kappa)/\sqrt{2}.
\end{equation}
As a result of using this formula, and the way we use the rotation operator $R(s)$, we can use the same schedule function when $A$ is not positive definite.

\subsection{Randomized method}\label{sec:RD}

A randomised QLSP algorithm inspired by AQC was first proposed in Ref.~\cite{PhysRevLett.122.060504} and further improved in recent work~\cite{jennings2023efficient}. 
The method uses the same $H_0$, $H_1$, and the same input state as the adiabatic quantum walk method, in order to approximate the desired eigenstate of $H_1$ with zero eigenvalue encoding the solution.

Let $H(v) = (1-f(v)) H_0 + f(v) H_1$ be the interpolating time-dependent Hamiltonian. 
Here we change the notation of the dimensionless time from $s$ to $v$ to avoid possible confusion, as in the Randomized method $v$ does not range from $0$ to $1$. 
Specifically, let 
\begin{equation}
    f(v) = \frac{-\kappa^2 \exp\left(-v\frac{\sqrt{\kappa^2+1}}{\sqrt{2}\kappa}\right) + \exp\left(v\frac{\sqrt{\kappa^2+1}}{\sqrt{2}\kappa}\right) + 2\kappa^2}{2(\kappa^2+1)}, 
\end{equation}
and 
\begin{align}
    v_a &= \frac{\sqrt{2}\kappa}{\sqrt{\kappa^2+1}} \log(\kappa\sqrt{1+\kappa^2} - \kappa^2),\\
    v_b &= \frac{\sqrt{2}\kappa}{\sqrt{\kappa^2+1}} \log(\sqrt{1+\kappa^2} + 1). 
\end{align}
Such a choice ensures that $f(v_a) = 0$ and $f(v_b) = 1$. 
The Randomized method then implements the operation 
\begin{equation}\label{eqn:RM_operator}
    e^{-i t_q H(v_q)} \cdots e^{-i t_2 H(v_2)} e^{-i t_1 H(v_1)}
\end{equation}
on the input state. 
Here $v_j = v_a + j(v_b-v_a)/q$, and $t_j$ are independent random variables. 

\begin{figure}[tbh]
    \centering
    \includegraphics[width=0.5\textwidth]{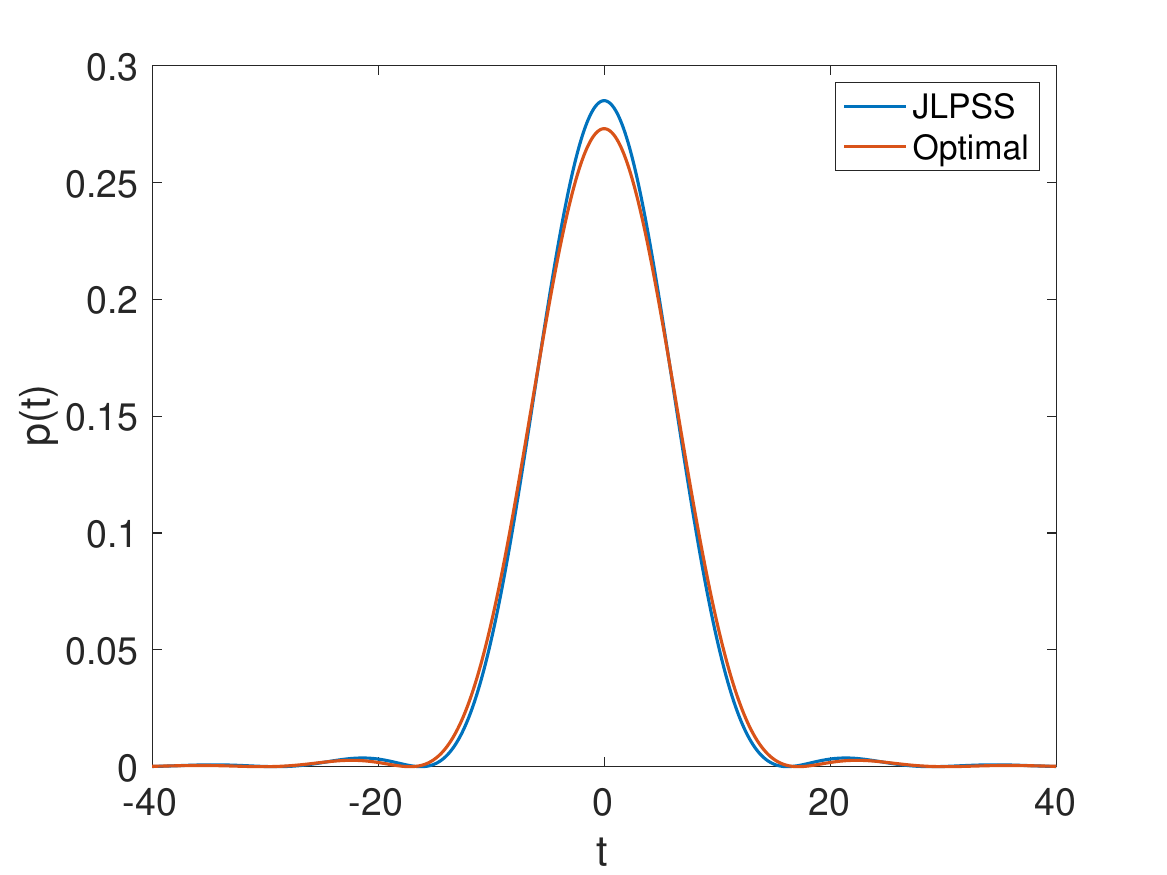}
    \caption{A comparison between two probability distribution functions used in the Randomized method. Here the legend ``JLPSS'' refers to the one in~\cite{jennings2023efficient} and ``Optimal'' to the one in~\cite{Sanders_2020}. The gap $\Delta(v_j)$ is chosen to be $0.5$ in this test.}
    \label{fig:RM_pdf_comp}
\end{figure}

In the original work~\cite{PhysRevLett.122.060504}, $t_j$ was chosen to be uniformly distributed in $[0, 2\pi/\Delta(v_j)]$. 
Recent work~\cite{jennings2023efficient} suggests to choose $t_j$ according to the probability density function 
\begin{equation}\label{eqn:RM_pdf_JLPSS}
    p_j(t) \propto \left( \frac{J_p(\Delta(v_j) |t|/2)}{\Delta^{p-1}(v_j) |t|^p } \right)^2, 
\end{equation}
where $J_p$ is a Bessel function of first kind and $p = 1.165$. 
It has the average time $\left<|t_j|\right> = 2.32132/\Delta(v_j)$. 
We remark that the way of constructing the optimal probability distribution has been proposed in~\cite{Sanders_2020}. 
It uses 
\begin{equation}\label{eqn:RM_pdf_optimal}
    p_j(t) = \left(\frac{1}{\sqrt{2\pi}} \sum_{l\geq 0} a_l \int_{-1}^1 \!\!\cos(xt\Delta(v_j)/2) (1-x^2) x^{2l} dx \right)^2\!\!, 
\end{equation}
where the coefficients $a_l$ can be numerically computed by minimising $\left< |t_j| \right>$. 
This optimal choice gives $\left< |t_j| \right> = 2.3160/\Delta(v_j)$, which is slightly better than~\cref{eqn:RM_pdf_JLPSS}. 
The corresponding probability distributions (\cref{eqn:RM_pdf_JLPSS} and~\cref{eqn:RM_pdf_optimal}) are almost the same (see~\cref{fig:RM_pdf_comp} for a comparison), so in this work we still use~\cref{eqn:RM_pdf_JLPSS}. 

Applying~\cref{eqn:RM_operator} to the input state produces the final state. 
It is shown in~\cite{jennings2023efficient} that, in order to achieve $1-\delta$ fidelity, it suffices to choose $q$ such that
\begin{equation}
    \left(1 - \frac{(v_b-v_a)^2}{q^2}\right)^q \geq 1 - \delta.
\end{equation}
Asymptotically $q \sim (v_b-v_a)^2/\delta \sim (\log \kappa)^2/\delta$. 
In practice, $\delta$ can be chosen to be $\mathcal{O}(1)$, and the state produced by the Randomized method will be passed into the filter to improve its fidelity. 

To implement the randomised evolution operator in \cref{eqn:RM_operator}, Ref.~\cite{jennings2023efficient} uses quantum signal processing (QSP) together with the Jacobi-Anger expansion. 
The cost of implementing an exponential $e^{-i t H(v)}$ is given as
\begin{equation}\label{eqn:Hsim_bound_JLPSS}
    3 \left( e \alpha_H |t|/2 + \log(2c/\gamma) \right)
\end{equation}
queries to the block encoding of $H(v)$, with $c = 1.47762$ and $\gamma$ the allowed error in each exponential. 
Therefore the overall number of queries to the block encoding of $H(v)$ becomes 
\begin{equation}
    3 \left( \frac{e \alpha_H}{2} \sum_{j=1}^q |t_j|  + q\log(2c/\gamma) \right). 
\end{equation}
That complexity analysis gives a non-asymptotic expression suitable for small times.
In contrast, Ref.~\cite{BabbushBerryNeven2019} gives an asymptotic analysis of the cost of implementing $e^{-i t H(v)}$ (with the same algorithm) to give the leading-order terms
\begin{equation}\label{eqn:Hsim_bound_BBN}
    2 \alpha_H |t| + 3^{2/3} (\alpha_H |t|)^{1/3} (\log(1/\gamma))^{2/3}
\end{equation}
for the number of queries to the block encoding of $H(v)$. 
The constant factor on $\alpha_H |t|$ is only $2$ here, which is better than the factor $3e/2$ in the bound of \cref{eqn:Hsim_bound_JLPSS}. 
Recent work~\cite{berry2024doubling} removes a factor of $2$ in the cost of quantum signal processing for Hamiltonian simulation.
That would remove a factor of 2 from both complexity estimates.
The leading-order term for the analysis of Ref.~\cite{BabbushBerryNeven2019} would then be $\alpha_H |t|$.
Therefore, with $\alpha_H=1$ the number of calls to the block encoding of $H(v)$ should be approximately (and no less than) $|t|$.
It is therefore reasonable to compare the total time in the Randomized method with the number of steps in the quantum walk approach.

\section{Numerical Tests}
\label{sec:Ne}

\begin{figure}[tbh]
    \centering
    \begin{tabular}{@{}p{0.8\linewidth}}
    \subfigimg[width=\linewidth]{(a)}{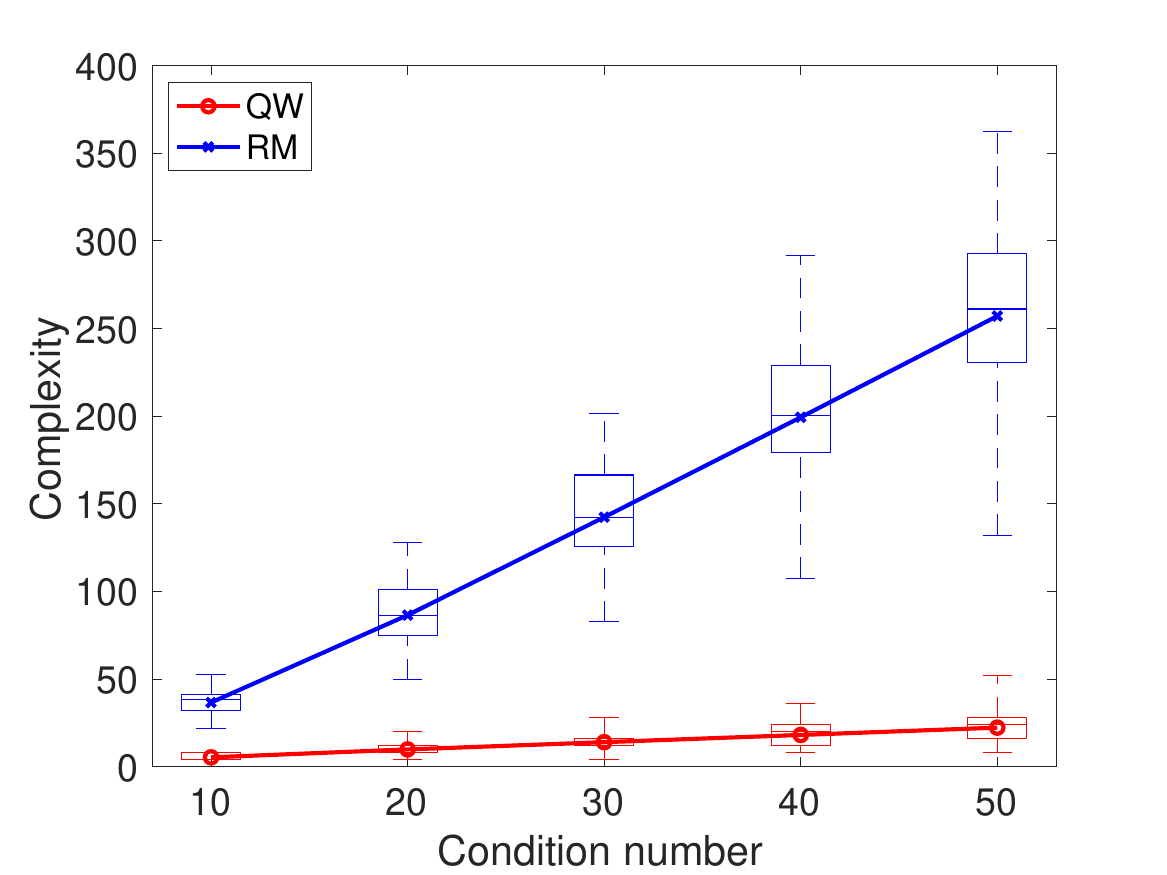} \\
    \subfigimg[width=\linewidth]{(b)}{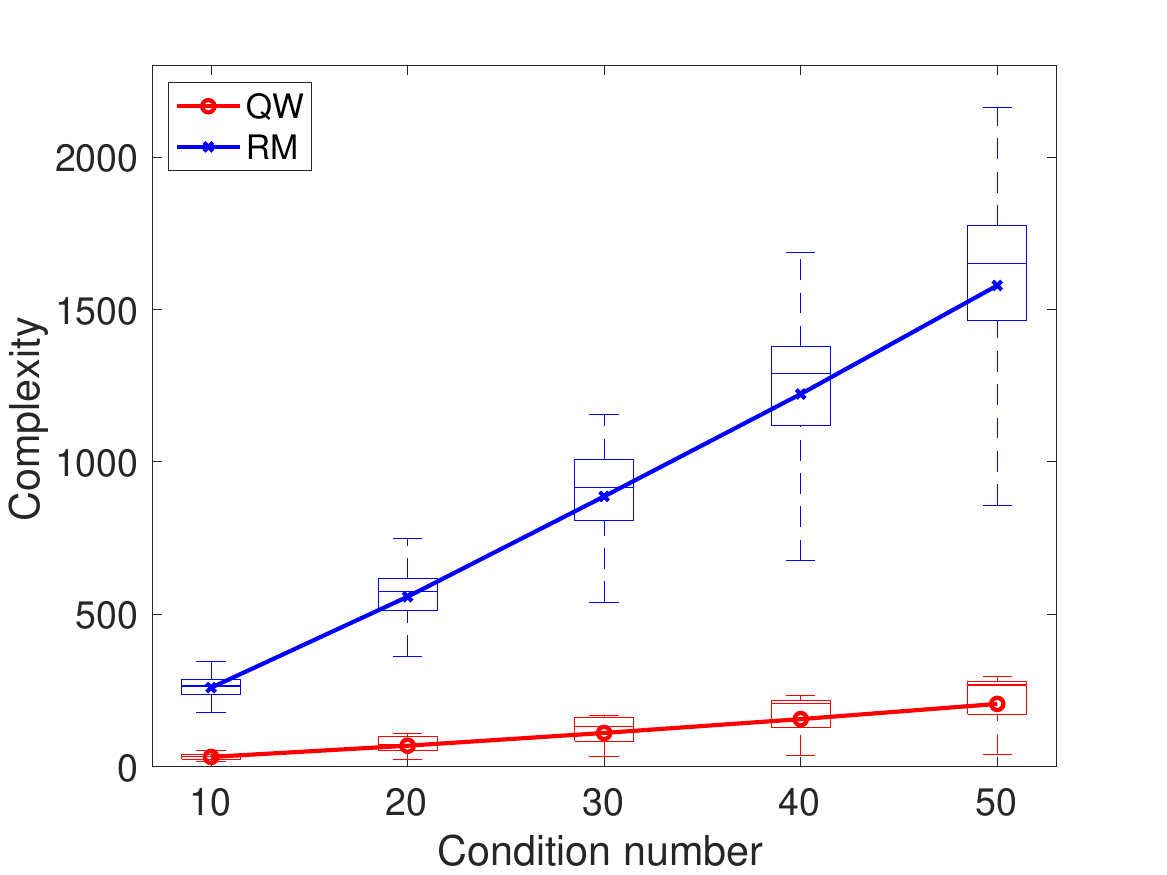}
    \end{tabular}
    \caption{Comparison between the quantum walk method and the Randomized method.
    The complexity is quantified by the number of calls to the block encoding of the matrix for the quantum walk case and by the total evolution time for the Randomized approach until we get the 2-norm error, $\Delta=0.4$ in the QW method and the infidelity error $\delta=0.4$ in for the Randomized method.
    The complexities are computed by the geometric mean of the 100 instances of $16$-dimensional matrices (solid line), and the box plots show their distributions. 
    Part (a): positive-definite case. Part (b): general non-Hermitian case. }
    \label{fig:comparison_geo_mean}
\end{figure}

Our main results are given in \cref{fig:comparison_geo_mean}, which includes the case where the matrix is $A$ positive definite (PD) and Hermitian as well as the case where $A$ is a general non-Hermitian matrix.
For the comparison we are quantifying the complexity for the discrete walk approach by the number of calls to the block encoding of $A$, whereas for the Randomized method we measure the cost by the total evolution time.
As discussed above, the total evolution time for the Randomized method enables us to lower bound the complexity in terms of block encodings, and thereby show that it is more costly than the discrete walk approach.

In \cref{fig:comparison_geo_mean} we can see an average complexity about $7$ times worse for the randomised method than the QW method.
This is comparing the number of calls to a block encoding of $A$ for the QW to a Hamiltonian evolution time for the RM.
In practice the simulation of the Hamiltonian evolution will incur extra overhead, so the advantage of the QW method will be even greater.

Moreover, we consider an improved probability distribution for the RM from Ref.~\cite{Sanders_2020}, so the performance of the RM as simulated is better than that presented in Ref.~\cite{jennings2023efficient}.
We also stress that the quantum walk performance is not only better on average over the thousands of tested instances of problems, but also for each instance tested.
With our results, we can avoid future misunderstanding of the performance of the method of Ref.~\cite{CostaAnYuvalEtAl2022}.

In both methods for solving the QLSP, we tested the cases where the matrices $A\in \mathbb{R}^{16\times 16}$ are positive definite and non-Hermitian. As shown in \cref{fig:comparison_geo_mean}, we consider matrices with a range of condition numbers $\kappa=10,20,30,40,50$, where for each condition number we generate $100$ independent samples.
We also consider matrices with lower dimensions in \cref{app:support}. 

The plots show the geometric mean of the complexity over the 100 instances for each condition number considered. We also included the error bars, with both the interquartile range and the complete data range.
The large size of the error bars results in part from the large value used as the threshold for the target error in the QLSP.
In particular, for the quantum walk, the principle is that we start with a small value for the total number of steps.
We then check the error between the actual and the ideal state.
If it is above the target error, we keep increasing the total number of steps until we reach this error.
However, the error between the ideal and the target state can vary rapidly with a small increment in the total number of steps, when we are not in the adiabatic regime.
Consequently, we can have large variation in complexity between different instances of the matrices.

\subsection{Quantum Walk method}

We start testing the AQC in the discrete framework for solving QLSPs.
For the positive-definite case the quantum circuit to construct the qubitised walk is shown in  \cref{fig:block} of \cref{app:block}.
The results for dimension $16\times 16$ are given in the first plot of \cref{fig:comparison_geo_mean}, and more detail and other matrix sizes are given in \cref{app:support}.
The results are given for 100 independent random matrices for each condition number.
We computed the norm of the difference between the ideal and the actual state, aiming for  $\|\ket{\tilde{x}} - \ket{x}\|\leq \Delta$ with $\Delta=0.4$ as the target error.
The complete QLSP algorithm uses an adiabatic component followed by filtering, both for the QW method and RM, and we denote the error for the adiabatic part of the algorithm as $\Delta$ to distinguish it from the final error $\epsilon$ obtained after filtering.
We use the schedule function as in Ref.~\cite{CostaAnYuvalEtAl2022} with $p=1.4$ (see \cref{sec:QW_meth}).
In the first plot of \cref{fig:comparison_geo_mean} (see also \cref{fig:AQC_PD} for more details and other dimensions) we observe a roughly linear growth of walk steps in condition number, as theoretically proven in \cite{CostaAnYuvalEtAl2022}. 

In the second plot of \cref{fig:comparison_geo_mean}, given with further details in \cref{fig:AQC_Nherm}, we present the results for non-Hermitian matrices.
See \cref{app:block} for explanation of the block encoding of the qubitised walk and the source code to generate these results is accessible at \cite{costa2024qlsp2}.
Similarly to the PD case, we generate 100 independent random matrices for each condition number.
In the non-Hermitian case, we again see the approximately linear growth of the walk steps in condition number. 

We also provide results in \cref{tab:Avg2} for non-Hermitian matrices, where rather than slowly increasing the total number of steps until we obtain the desired threshold, we use the same number of total steps for all instances.
The total number is chosen to obtain root-mean-square (RMS) error (for 100 instances) as close as possible to $\Delta=0.4$ but never higher.
We also include a table for PD matrices in \cref{app:support}.
From the results in \cref{tab:Avg2}, we can infer the constant factor $\alpha$ in $T=\alpha \kappa /\Delta$ which we will find in practice for the QLSP. In particular, for $\kappa=50$, we find $\alpha=1.84$, which gives a constant factor about 1,250 smaller than what is presented in Ref.~\cite{CostaAnYuvalEtAl2022}.

In addition to the randomly tested matrices without a specific application, we included two matrices from the collection presented in \cite{sparse_matrix_collection}, which provides a variety of sparse matrices used for benchmarking across different applications. The results for these practical cases are reported in \cref{tab:merged_practical}.

\begin{table*}[t]
\centering
\begin{tabular}{|c|c|c|c|c|c|}
\hline
\multirow{2}{*}{$\kappa$} & \multicolumn{2}{c|}{Quantum Walk} & \multicolumn{2}{c|}{Randomized Method}&
\multirow{2}{*}{Ratio}\\
\cline{2-5}
 & Number of Steps & $\Delta$ & Averaged Time & $\Delta$ & \\
\hline
$10$ & $36$  & $0.348$ & $281$  & $0.395$ & $7.81$\\
$20$ & $76$  & $0.381$ & $604$  & $0.397$ & $7.95$\\
$30$ & $120$ & $ 0.400$ & $963$  & $0.398$ & $8.03$\\
$40$ & $176$ & $0.399$ & $1330$ & $0.397$ & $7.56$\\
$50$ & $232$ & $0.397$ & $1722$ & $0.399$ & $7.42$\\
\hline
\end{tabular}
\caption{Comparison between the Quantum Walk and Randomized Methods for a range of condition numbers applied to non-Hermitian matrices of dimension $16 \times 16$. For the discrete adiabatic method, the RMS error, measured as the norm of the difference in quantum states, is computed over 100 instances for each condition number. For the randomized method, the average evolution time required to achieve an average error $\Delta$ of approximately $0.4$ is reported. The RMS error is also computed over 100 instances, with 200 repetitions per instance.
The Ratio column shows the factor of improvement of the QW over RM.
}
\label{tab:Avg2}
\end{table*}

\begin{table*}[t]
\centering
\begin{tabular}{|c|c|c|c|c|c|c|}
\hline
\multirow{2}{*}{Matrix } & \multirow{2}{*}{$\kappa$} 
& \multicolumn{2}{c|}{Quantum Walk} 
& \multicolumn{2}{c|}{Randomized Method} 
& \multirow{2}{*}{Ratio } \\
\cline{3-6}
& & Number of Steps & $\Delta$ & Averaged Time & $\Delta$ &\\
\hline
Directed Graph – ID=168 & $4.041 \times 10^2$ 
&$2.416\times 10^3$ & $0.398$ 
& $2.315 \times 10^4$ & $0.406$ & $9.58$\\
Circuit Simulation Problem – ID=1199 & $6.302\times 10^5$ 
&$3.980\times 10^4$ & $0.399$ 
& $1.817 \times 10^7$ & $0.402$ & $457$\\
\hline
\end{tabular}
\caption{Performance comparison between the Quantum Walk  and RM for two non-Hermitian matrices of dimension $32 \times 32$ from \cite{sparse_matrix_collection}. For each matrix, we report its application, ID, condition number, and performance metrics including number of steps and error for QW, and averaged time and error for RM.
The Ratio column shows the factor of improvement of the QW over RM.}
\label{tab:merged_practical}
\end{table*}

\subsection{Randomized method}

In the Randomized method, we measure the cost by the total time, defined as $\sum_{j=1}^q |t_j|$, to achieve desired accuracy.
Hamiltonian evolution may be simulated using calls to block encodings of the Hamiltonian via quantum signal processing.
With $\|H\|=1$, the simulation via the most efficient known scheme requires at least as many calls to the Hamiltonian as the time \cite{berry2024doubling}, though the cost can be larger (as a ratio to time) for shorter simulation times, as analysed in Ref.~\cite{jennings2023efficient}.
It is therefore reasonable to compare the total time here to the number of walk steps;
it enables us to lower bound the true cost of the RM, to show that it is more costly than the QW.
See \cref{sec:RD} for more details, and the source code to generate these results are accessible at \cite{costa2024qlsp}.

In our tests, we record the averaged minimal total time to reduce the RMS error below $0.4$ over the 100 independent samples for each dimension and condition number, with the same sample matrices as for the QW method.
For each sample we perform 200 repetitions with new randomly selected times.
Such a choice of RMS error is comparable to the error threshold in the QW tests. 
We remark that the actual numerical RMS error is slightly smaller than $0.4$, since in the Randomized method only the number of exponentials $q$ is tunable and the resulting errors are changing discontinuously. 
As an illustration, results for $16$-dimensional non-Hermitian matrices are shown in~\cref{tab:Avg}, and more results in PD cases can be found in~\cref{app:support}. 

From \cref{fig:comparison_geo_mean}, which can be seen with further details in \cref{fig:RM_PD,fig:RM_nonH}, we observe roughly linear growth of time in condition number, as theoretically proven. 
We also observe that the total time for the non-Hermitian case is about $6.4$ times larger than that for the PD case. 
In both PD and non-Hermitian cases, the overall complexity of the RM is significantly higher than that of the QW method. 
Because we are comparing evolution time for the RM to the number of steps for the QW, the advantage will be even greater when the complexity of implementing the time evolution is taken into account.

Same as the QW method, we also perform a numerical test of RM using the same two practical matrices from~\cite{sparse_matrix_collection}. The results are reported in~\cref{tab:merged_practical} and show that RM can be $9.2$ times worse than QW in the directed graph example, and $4.0$ times worse in the circuit simulation problem.

\section{Cost with filter}
\label{sec:filter}

Typically, we would aim to perform the discrete adiabatic algorithm followed by a filtering step to ensure the correct eigenvalue has been obtained; this filtering procedure is a common step in both the discrete adiabatic and randomized methods, and incurs similar cost in each case. First assume that the adiabatic algorithm gives an infidelity error $\delta$, in the sense that the probability of measuring the correct eigenstate is $1-\delta$.
The relation between $\Delta$ (2-norm error) and infidelity error $\delta$ is $\delta = \Delta^2 - \Delta^4/4$.
Then if we were to perform filtering as described in \cite{CostaAnYuvalEtAl2022} (see Eq.~(113) of that work), the complexity of the filter would be $\kappa\ln(2/\epsilon_f)$, with $\epsilon_f$ the parameter for the filter.
In that work it is shown that the 2-norm error at the end is upper bounded by $\epsilon_f$ (for $\delta>1/2$), and $\epsilon_f^2$ is also the maximum probability for failure (accepting an incorrect state as correct).

If the state fails the filtering step, then the adiabatic algorithm and filter are repeated.
It should also be taken into account that the error probability will be increased if there are multiple failures and filters.
The set of possible outcomes is as follows.
\begin{enumerate}
    \item With probability $1-\delta$ the correct state is obtained the first time, which then passes the filtering test.
    \item With probability $\delta$ there is an incorrect state. With probability no greater than $\delta\epsilon_f^2$ this state will be accepted regardless, and otherwise it will be flagged as a failure.
    The probability of this occurring is at least $\delta(1-\epsilon_f^2)$.
    \item To upper bound the probability of error we should take the probability of performing the procedure a second time as $\delta(1-\epsilon_f^2)$.
    The probability of incorrectly accepting the wrong state is then upper bounded as $\delta^2\epsilon_f^2(1-\epsilon_f^2)$.
\end{enumerate}
In general, the probability of flagging a failure and repeating the measurement $n$ times is $[\delta(1-\epsilon_f^2)]^n$.
For the average number of repetitions, it will be, in this worst-case scenario for error,
\begin{equation}
    \sum_{n=0}^\infty (n+1)[\delta(1-\epsilon_f^2)]^n = \frac{1}{[1-\delta(1-\epsilon_f^2)]^2}.
\end{equation}

For the costing, we need to bear in mind that for the cases where the filter fails the LCU approach will on average flag a failure after half the maximum number of operations.
This means that the cost of the filter can be given as
\begin{align}
    &\kappa\ln(2/\epsilon_f) + \frac 12 \kappa\ln(2/\epsilon_f) \left( \frac{1}{[1-\delta(1-\epsilon_f^2)]^2}-1 \right)\nn 
    &= \frac 12 \kappa\ln(2/\epsilon_f) \left( \frac{1}{[1-\delta(1-\epsilon_f^2)]^2}+1 \right)
\end{align}

If the number of steps in the adiabatic algorithm to obtain error $\Delta$ is $\alpha\kappa/\Delta$, then the cost of the adiabatic repetitions is
\begin{equation}
    \frac{\alpha\kappa}{\Delta[1-\delta(1-\epsilon_f^2)]^2} \, ,
\end{equation}
for a total complexity
\begin{equation}\label{eq:complex}
    \frac 12 \kappa\ln(2/\epsilon_f) \left( \frac 1{[1-\delta(1-\epsilon_f^2)]^2}+1 \right) + \frac{\alpha\kappa}{\Delta[1-\delta(1-\epsilon_f^2)]^2} \, .
\end{equation}
Here we can use $\delta = \Delta^2 - \Delta^4/4$ or $\Delta=\sqrt{2-2\sqrt{1-\delta}}$ to give the expression in terms of a single variable ($\delta$ or $\Delta$).
The value of $\kappa$ factors out, and given $\alpha$ one can optimise $\delta$ or $\Delta$ for a given $\epsilon_f$ to minimise the cost.
In the case where the cost scales with $1/\delta$ instead of $1/\Delta$, then the expression for the total complexity is changed slightly to
\begin{equation}\label{eq:complex2}
    \frac 12 \kappa\ln(2/\epsilon_f) \left( \frac 1{[1-\delta(1-\epsilon_f^2)]^2}+1 \right) + \frac{\beta\kappa}{\delta[1-\delta(1-\epsilon_f^2)]^2} \, ,
\end{equation}
where we are using $\beta$ for the constant factor,
and a similar minimisation is possible.
It is important to stress that, although the filtering step is in common between the methods, its cost depends on $\Delta$.
Because the different methods have different optimal values of $\Delta$, the cost of the filtering step is different between the methods.

A further adjustment to the complexity can be made by noting that $\epsilon_f$ here is just an upper bound on the 2-norm error $\epsilon$.
As per Eq.~(J3) of \cite{CostaAnYuvalEtAl2022}, the fidelity in the filtered state can be lower bounded as
\begin{equation}
    f \ge \frac{1-\delta}{1-\delta+\delta\epsilon_f^2} .
\end{equation}
Then the 2-norm error is $\epsilon=\sqrt{2-2\sqrt{f}}$, so solving for $\epsilon_f$ gives
\begin{equation}
    \epsilon_f = \epsilon \frac{\sqrt{(1-\delta)(1-\epsilon^2/4)}}{(1-\epsilon^2/2)\sqrt\delta} .
\end{equation}
We are using the notation $\epsilon_f$ for the parameter used in the filter, and $\epsilon$ for the final allowable error.
We can use this expression for $\epsilon_f$ in Eq.~\eqref{eq:complex} or Eq.~\eqref{eq:complex2} to give a slightly tighter bound for the complexity.

\begin{figure}[tbh]
    \centering
    \includegraphics[width = 0.45\textwidth]{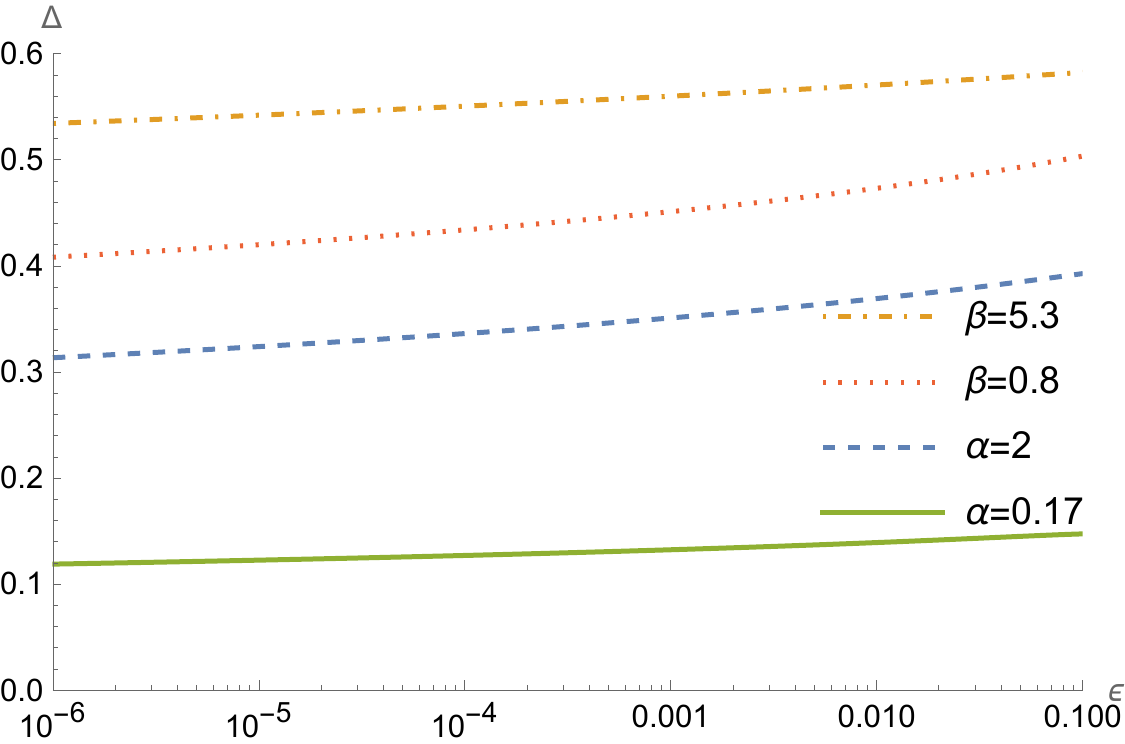}    
    \caption{The values of $\Delta$ as a function of $\epsilon$ that minimise the cost.
    The results are for $\alpha=0.17$ (green) and $\alpha=2$ (blue) using the cost in Eq.~\eqref{eq:complex}.
    Results using Eq.~\eqref{eq:complex2} are shown for $\beta=0.8$ (red) and $\beta=5.3$ (orange).
    }
    \label{fig:optdelt}
\end{figure}

The results for numerically solving for the optimal $\Delta$ are shown in Fig.~\ref{fig:optdelt} for a range of values of $\alpha$.
For typical values of $\alpha$, optimal values of $\Delta$ are around $0.4$.
When $\alpha$ is reduced to $0.17$ (as for the PD case for the quantum walk), the optimal $\Delta$ is around $0.15$.
For the Randomized method with cost scaling as $\beta/\delta$ instead, the optimal values of $\delta$ are around $0.5$.
In the preceding section, we have used the same $\Delta$ for the different methods despite the different optimal values to give a fair comparison of the approaches.

In this result it is assumed that the scaling holds exactly, even for larger values of $\Delta$.
In reality, this scaling only holds in the asymptotic limit of small $\Delta$, and determining the scaling constant with larger values of $\Delta$ can give larger values.
In Table \ref{tab:multieps} we show the results for each case calculated with a range of values of $\Delta$.
In the case of the QW with general matrices, the performance is notably better for smaller values of $\Delta$, and we estimate $\alpha=1.84$, $1.66$ and $1.37$, for $\Delta\approx 0.4$, $0.3$ and $0.2$,  respectively.
To more accurately estimate the optimal values of $\Delta$, in each case we have used the results for three values of $\Delta$ (those shown in Table \ref{tab:multieps}) to interpolate the scaling constants $\alpha$ and $\beta$ as functions of $\Delta$.

\begin{figure}[tbh]
    \centering
    \includegraphics[width = 0.45\textwidth]{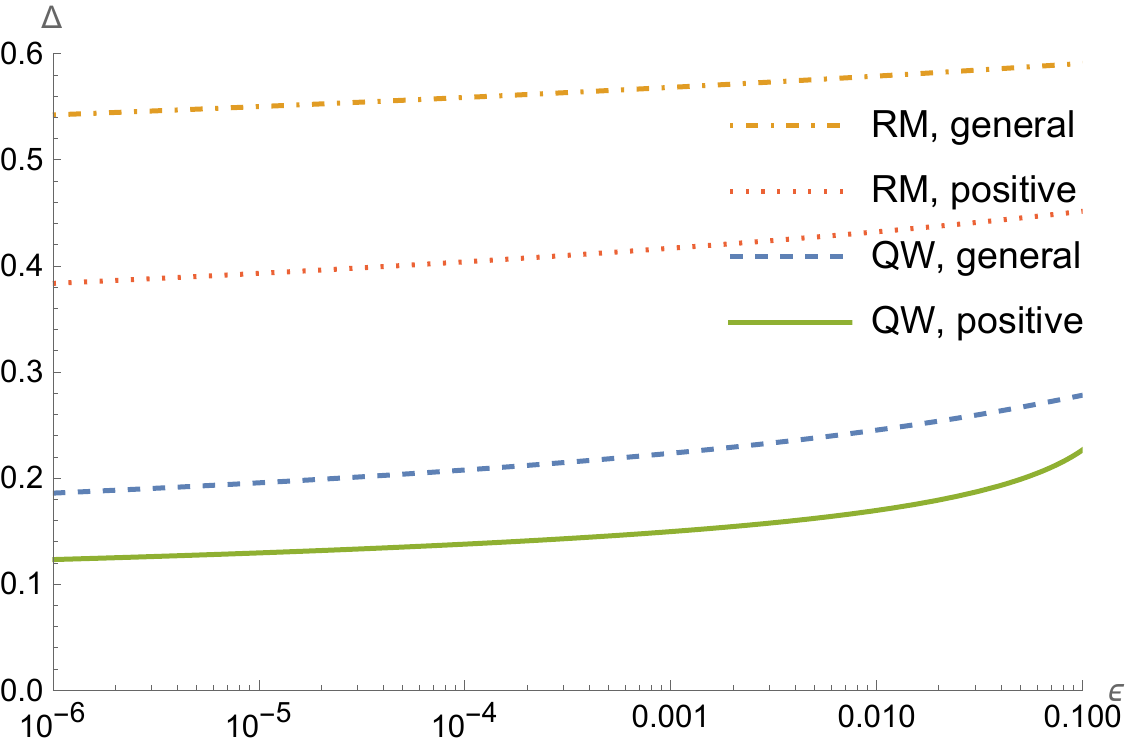}    
    \caption{The values of $\Delta$ as a function of $\epsilon$ that minimise the cost.
    We include results assuming various values of $\alpha$, as well as with interpolation of the value of $\alpha$ as a function of $\Delta$.
    }
    \label{fig:optdelt2}
\end{figure}

The resulting optimal values of $\Delta$ are shown in Fig.~\ref{fig:optdelt2}.
In the case of the QW with general matrices, the predicted optimal values of $\Delta$ are reduced by about $0.1$ as compared to those in Fig.~\ref{fig:optdelt}.
There are more minor differences in the other cases.
The value $\Delta=0.4$ used in Section \ref{sec:Ne} is in the middle of the range for the optimal values, which is the reason it was chosen as a common value for the numerical comparisons above.

\begin{table*}[tbh]
\centering
\begin{tabular}{|c|c|c|c|c|c|c|}
\hline
\multirow{2}{*}{Case} & \multicolumn{6}{c|}{Target $\Delta$} \\
 \cline{2-7}
& 0.1 & 0.2 & 0.3 & 0.4 & 0.5 & 0.6 \\
\hline
QW, positive & 108 (0.0971) & 56 (0.1990) & 32 (0.2821) & & & \\
QW, general & & 344 (0.1986) & 288 (0.2878) & 232 (0.3967) & & \\
RM, positive & & & 421 (0.3012) & 271 (0.3916) & 195 (0.4959) & \\
RM, general & & & & 1722 (0.3991) & 1100 (0.4987) & 771 (0.5989)\\
\hline
\end{tabular}
\caption{The number of steps or averaged time for a range of values of $\Delta$ that are used for the interpolation.
We show both the QW and RM, both for positive definite and general matrices, with $16\times 16$ matrices and $\kappa=50$ for all.
The values in brackets are the actual values of $\Delta$ obtained.
}
\label{tab:multieps}
\end{table*}

\begin{figure}[tbh]
    \centering
    \includegraphics[width = 0.45\textwidth]{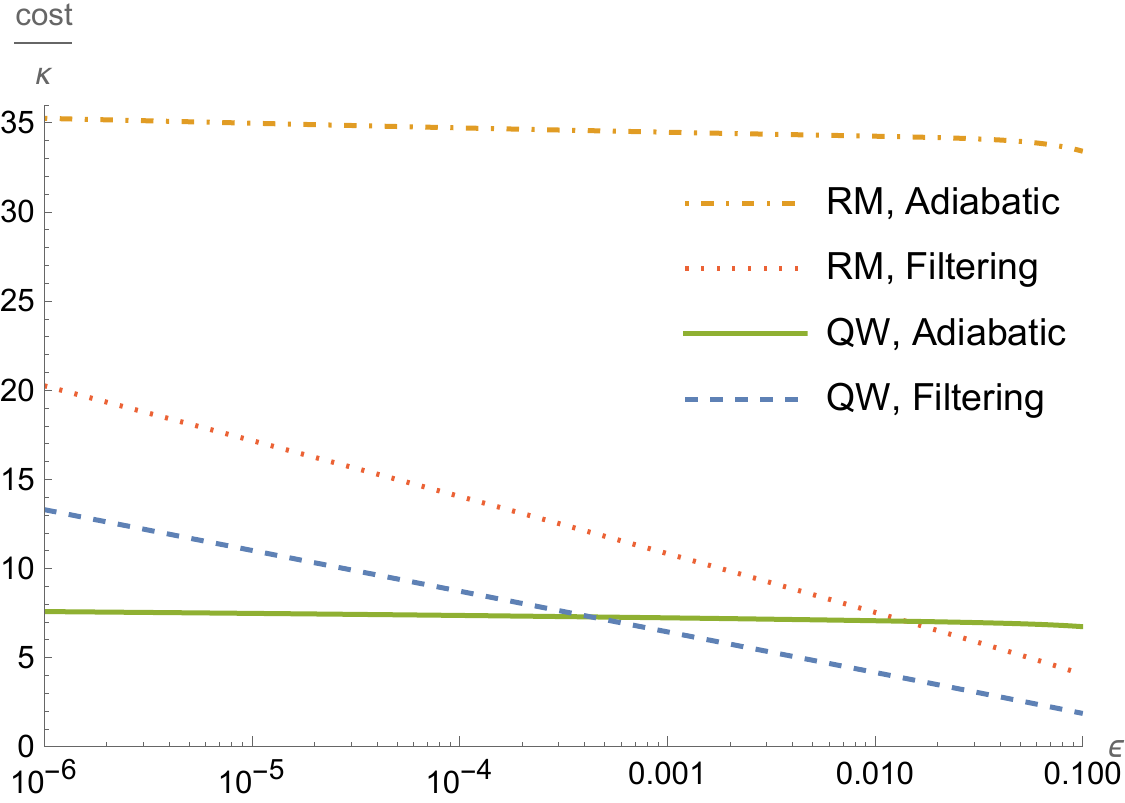}
    \caption{The two contributions to the minimised costs as a function of $\epsilon$ for the discrete adiabatic approach and for the randomised walk.
    In both cases the scaling constant is fitted as a function of $\Delta$.}
    \label{fig:fourcosts}
\end{figure}

The division of the costs between the adiabatic and filtering steps is shown in Fig.~\ref{fig:fourcosts} for the {interpolated values of $\alpha$ for the PD case using the quantum walk.
The contribution to the cost from the adiabatic part of the algorithm is relatively independent of $\epsilon$, whereas the filtering has a clear $\ln(1/\epsilon)$ dependence as would be expected.
For values of $\epsilon$ above about $0.0004$, the majority of the contribution to the cost is from the adiabatic part of the algorithm.
For smaller values of $\epsilon$ the cost of the filtering step is larger, but these are smaller values of $\epsilon$ than would be expected in applications.
Because QLSP outputs the solution encoded in a quantum state, classical results can only be obtained from sampling.
It would require millions of samples to provide an accuracy better than $0.0004$, and there would not be an advantage to solving the QLSP to significantly higher accuracy than the sampling.

The Randomized method with $\beta$ interpolated is also shown in Fig.~\ref{fig:fourcosts}.
The adiabatic part of the algorithm is significantly more costly than the filtering part for the entire range of $\epsilon$ shown.
It takes unreasonably small values of $\epsilon$ below about $10^{-12}$ for the filtering to be more costly.
Both the adiabatic and filtering parts of the algorithm are more costly than for the quantum walk, with the adiabatic part being much more costly.
This means that the \emph{total} cost (including both the adiabatic component and filtering component) for the randomised walk is substantially larger than for the discrete adiabatic approach, as shown in Fig.~\ref{fig:bothcosts}.

\begin{figure}[tbh]
    \centering
    \includegraphics[width = 0.45\textwidth]{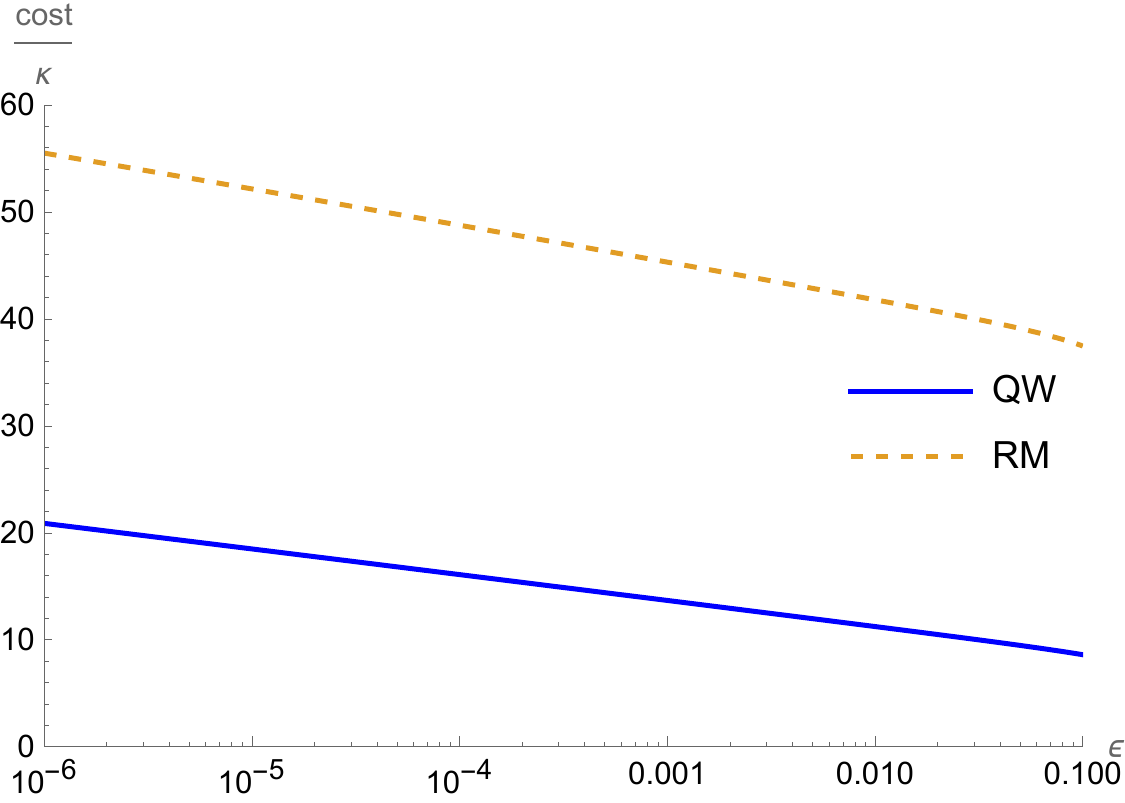}
    \caption{The total costs as a function of $\epsilon$ for the discrete adiabatic approach with $\alpha$ fitted as a function of $\Delta$ (blue) and the randomised walk with $\beta$ fitted as a function of $\delta$ (orange).}
    \label{fig:bothcosts}
\end{figure}

\section{Conclusion}\label{sec:conc}

The optimal quantum linear systems algorithm of \cite{CostaAnYuvalEtAl2022} works by first applying a quantum walk to provide a result with low accuracy, then filtering the resulting state to improve the precision to the desired level.
The initial quantum walk has an upper bound on its error with a large constant factor in \cite{CostaAnYuvalEtAl2022} which would suggest it is a costly algorithm despite having optimal asymptotic complexity.

Here we have numerically tested this initial quantum walk with random matrices and found that the constant factor is about 1,200 times smaller.
This means that the algorithm will be highly efficient in practice, with the cost of the quantum walk comparable to the cost of the filtering. The cost of the quantum walk is typically only a little larger than the cost of the filtering, but in special cases, such as positive-definite matrices or $\epsilon \ll 1$, its cost is even smaller, as would be expected since the filtering step has the factor of $\log(1/\epsilon)$ in the complexity.

Reference \cite{jennings2023efficient} provided a tighter upper bound on the complexity for an approach with randomised evolution times for initial adiabatic evolution, followed by a filtering step that is similar to \cite{CostaAnYuvalEtAl2022}.
In our numerical testing of that approach, we find that it is about an order of magnitude more costly in practice, despite having a tighter analytic bound on the complexity,  where we have $1/\Delta$ in the cost of the adiabatic part vs $1/\delta$ in the cost of the randomized method.
It is somewhat ambiguous to compare the complexity because it is in terms of an evolution time rather than the number of steps of a quantum walk.
The evolution time is about 7 times the number of steps of the quantum walk needed for the discrete adiabatic approach, and in practice the cost to simulate that Hamiltonian evolution would be even larger.

\subsection*{Acknowledgements}

DWB worked on this project under a sponsored research agreement with Google Quantum AI. DWB is also supported by Australian Research Council Discovery Projects DP210101367 and DP220101602. 
DA acknowledges the support by the Department of Defense through the Hartree Postdoctoral Fellowship at QuICS, and the seed grant at the NSF Quantum Leap Challenge Institute for Robust Quantum Simulation (QLCI grant OMA-2120757).

\appendix
\section{Lower bound}
\label{app:lower}

The $\Omega(\kappa\log(1/\epsilon))$ part of the lower bound from \cite{RobinAram} can be obtained via a minor modification of the proof from HHL \cite{Harrow_2009}.
Adopting the notation of \cite{Harrow_2009}, in Eq.~(4) of that work a unitary $U$ is constructed such that for $T+1\le t\le 2T$, applying $U^t$ to $\ket{1}\ket{\psi}$ yields $\ket{t+1}\otimes U_T\cdots U_1\ket{\psi}$ for some sequence of unitary operators $U_j$.
Now define
$A= I-U \epsilon^{1/T}$
so the condition number $\kappa = \mathcal{O}(T/\log(1/\epsilon))$ and we can expand
\begin{equation}
A^{-1} = \sum_{k=0}^\infty U^k \epsilon^{k/T}.
\end{equation}
After applying $A^{-1}$ one would obtain $t$ in the range $T+1$ to $2T$ with $\Theta(\epsilon)$ probability.
Therefore, solving the linear system with precision $\epsilon$ provides an unbounded-error algorithm for implementing $T$ steps $U_T\ldots U_1$.

It is straightforward to choose the operations $U_j$ so that implementing them in succession gives the parity of $T$ bits, and each corresponds to a single query to these bits.
An unbounded-error algorithm for determining the parity of $T$ bits requires $\Omega(T)$ queries \cite{parity1,parity2}.
That means there must be a lower bound $\Omega(T)$ for the number of queries to the operators $U_j$.
Now $A$ can be block encoded with a single query to $U_j$, there is similarly a lower bound $\Omega(T)$ to the number of calls to the block encoding of $A$.
Since $\kappa=\mathcal{O}(T/\log(1/\epsilon))$, we have $T=\Omega(\kappa\log(1/\epsilon))$, so the complexity of the linear systems algorithm is $\Omega(\kappa\log(1/\epsilon))$.

\section{Block encoding}
\label{app:block}

\subsection{Positive definite $A$}
Now we show how to block encoding $H(s)$ from \cref{eq:Ham_prob}, when we have $H_0$ and $H_1$ given by \cref{eq:H0} and \cref{eq:H1} respectively.
The general principle for the block encoding of our Hamiltonian is to block encode both $H_0$ and $H_1$, and apply the following controlled operation
\begin{equation}
\label{eq:Sel_op}
\textsc{sel} = \ket{0}\bra{0} \otimes U_{H_0} + \ket{1}\bra{1} \otimes U_{H_1},
\end{equation}
to select between $H_0$ and $H_1$.
To give the linear combination of $H_0$ and $H_1$ we apply the one-qubit rotation on the ancilla qubit
\begin{equation}
\label{eq:C-rot}
R(s)=\frac{1}{\sqrt{\left(1-f(s)\right)^2+f(s)^2}}\begin{pmatrix} 1-f(s) &  f(s)\\
 f(s) & -(1-f(s))
\end{pmatrix}.
\end{equation}
In the following we describe the details of this implementation.

We denote the unitary for the block encoding of $A$ as $U_A$, which acts on an ancilla denoted with subscript $a$ and the system such that
\begin{equation}
{}_a \! \bra{0} U_A \ket{0}_a = A.
\end{equation}
We also denote the unitary oracle for preparing $\ket{b}$ as $U_b$ such that
\begin{equation}
\label{eq:Oracb}
U_b \ket{0} = \ket{b}.
\end{equation}
As well as the ancilla system used for the block encoding of $A$, we use three ancilla qubits. These ancilla qubits are used as follows.
\begin{enumerate}
    \item The first selects between the blocks in $H_0$ and $H_1$.
    \item The next is used for preparing the combination of $H_0$ and $H_1$.
    \item The third is used in implementing $Q_b$.
\end{enumerate}
We use the subscripts $a_h$, $a_1$ and $a_2$ for these three qubits.
Note that although we have called $a_h$ an ancilla qubit, it part of the systam that the Hamiltonian acts upon, not an ancilla used in the block encoding.
This means that we do not apply the reflection operator on that ancilla to construct the walk operator.

To make the block encoding $H(s)$ simpler let us rewrite \cref{eq:H1} as
\begin{equation}
H_1 = \begin{pmatrix} I &  0\\
 0 & Q_b
\end{pmatrix}\begin{pmatrix} 0 &  I\\
 I & 0
\end{pmatrix}\begin{pmatrix} A &  0\\
 0 & A
\end{pmatrix}\begin{pmatrix} I &  0\\
 0 & Q_b
\end{pmatrix},
\end{equation}
and  \cref{eq:H0} as
\begin{equation}
H_0 = \begin{pmatrix} I &  0\\
 0 & Q_b
\end{pmatrix}\begin{pmatrix} 0 &  I\\
 I & 0
\end{pmatrix}\begin{pmatrix} I &  0\\
 0 & I
\end{pmatrix}\begin{pmatrix} I &  0\\
 0 & Q_b
\end{pmatrix}.
\end{equation}
This new way to express the Hamiltonians will then facilitate the construction of the select operator  since the total Hamiltonian can be expressed as
\begin{align}
\label{eq:Hs}
H(s)&= \begin{pmatrix} I &  0\\
 0 & Q_b
\end{pmatrix}\left[(1-f(s))\begin{pmatrix} 0 &  I\\
 I & 0
\end{pmatrix}\right.\nn 
&\quad+ \left. f(s)\begin{pmatrix} 0 &  I\\
 I & 0
\end{pmatrix}\begin{pmatrix} A &  0\\
 0 & A
\end{pmatrix}\right]\begin{pmatrix} I &  0\\
 0 & Q_b
\end{pmatrix}.
\end{align}
For the first block matrix inside the bracket in the equation above, we have
\begin{equation}
\label{eq:X}
\begin{pmatrix} 0 &  I\\
 I & 0
\end{pmatrix} = \sigma^x_{a_h}\otimes I \, ,    
\end{equation}
where $\sigma^x_{a_h}$ indicates that we have the Pauli $X$ operator acting on the register $a_h$. Now, for the other block matrix which comes from  $H_1$, we have
\begin{equation}
\label{eq:A}
\begin{pmatrix} A &  0\\
 0 & A
\end{pmatrix} = I_{a_h}\otimes A \, ,  
\end{equation}
where we have included $I_{a_h}$ to indicate that the operation acts as the identity on the first qubit.
Moreover, we have $U_A$ as the unitary for the block encoding of $A$, with ancilla qubit $a$.
Thus when we combine \cref{eq:X} and \cref{eq:A}, the two block matrices for $H_1$, we have
\begin{align}
 &\left(\sigma^x_{a_h}\otimes I_N \otimes I_a\right)\left(I_{a_h}\otimes U_A\right)\nn 
 &=\ket{0}\bra{1}_{a_h}\otimes U_A + \ket{1}\bra{0}_{a_h}\otimes U_A \, .      
\end{align}
However, the encoding above does not guarantee the hermiticity of $H_1$, which can be fixed by the following expression
\begin{equation}
\label{eq:middle_H1}
\ket{0}\bra{1}_{a_h}\otimes U_A + \ket{1}\bra{0}_{a_h}\otimes U_A^{\dagger} \, .      
\end{equation}
Therefore, we have \cref{eq:middle_H1} as the block encoding of the two middle operators of $H_1$ and $\sigma^x_{a_h}\otimes I_N \otimes I_a$ as the block encoding of the two middle operators of $H_0$.
We can therefore select between $H_0$ and $H_1$ using the operator
%From what is given in \cref{eq:Hs} we can, thus, make use of these two block encodings to build the select operator \cref{eq:Sel_op}, which we denote $U_{A(f)}$, and only after the remaining part which is the same for both, 
\begin{align}
\label{eq:sel_2}
U_{A(f)} &= \ket{0}\bra{0}_{a_1} \otimes\sigma^x_{a_h}\otimes I_N \otimes I_a\nn 
&\quad+ \ket{1}\bra{1}_{a_1} \otimes \left(\ket{0}\bra{1}_{a_h}\otimes U_A + \ket{1}\bra{0}_{a_h}\otimes U_A^{\dagger}\right) ,
\end{align}
together with the operators involving $Q_b$ in \cref{eq:Hs} and rotation on the ancilla qubit $a_1$.
The operator involving $Q_b$ in \cref{eq:Hs} can be rewritten as
\begin{align}
\begin{pmatrix} I &  0\\
 0 & Q_b
\end{pmatrix} &= \ket{0}\bra{0}_{a_h}\otimes I_N + \ket{1}\bra{1}_{a_h}\otimes Q_b\nonumber\\
&= \ket{0}\bra{0}_{a_h}\otimes I_N + \ket{1}\bra{1}_{a_h}\otimes \left(I_N -\ket{b}\bra{b}\right)\nonumber\\
&=\ket{0}\bra{0}_{a_h}\otimes I_N\nn 
&\quad+ \frac{1}{2}\ket{1}\bra{1}_{a_h}\otimes \left(I_N +\left(I_N -2\ket{b}\bra{b}\right)\right)\nonumber\\
&=\ket{0}\bra{0}_{a_h}\otimes I_N\nn 
&\quad+ \frac{1}{2}\ket{1}\bra{1}_{a_h}\otimes \left[I_N +\left(I_N -2U_b\ket{0}\bra{0}U_b^{\dagger}\right)\right].
\end{align}
In the second equality, we have written the projection in terms of a sum of the identity and a projector, and in the last equality, we used the oracle for preparing $\ket{b}$ as given in \cref{eq:Oracb}. We can block encode this projector using the ancilla $a_2$. We simply need to create this ancilla in an equal superposition and use the following unitary operation
\begin{align}
CU_{Q_b} &= \left(I_{a_h}\otimes I_{a_2}\otimes U_b^{\dagger} \right)\bigg\{\ket{0}\bra{0}_{a_h}\otimes I_{a_2}\otimes I_N\nn 
&\quad + \ket{0}\bra{0}_{a_h}\otimes \left[\ket{0}\bra{0}_{a_2}\otimes I_N + \ket{1}\bra{1}_{a_2}\right.\nn 
&\quad\otimes \left.\left(I_N -2\ket{0}\bra{0}\right)\right]\bigg\}\left(I_{a_h}\otimes I_{a_2}\otimes U_b \right).   
\end{align}
To block encode $H(s)$ as in \cref{eq:Hs}, the block encoding above is applied before and after the select operator $U_{A(f)}$.

We also need to perform the rotation $R(s)$ in order to interpolate between the Hamiltonians. The interpolation operator can then act on the ancilla qubit used in the select operator. However, instead of applying the inverse of $R(s)$, we simply perform a Hadamard and  to produce the right weights for the Hamiltonians we need to apply the rotation operator controlled by ancilla qubit $a_h$ as
\begin{equation}
CR(s)= \ket{0}\bra{0}_{a_h}\otimes R(s)_{a_1} +   \ket{1}\bra{1}_{a_h}\otimes \mathcal{H}_{a_1} 
\end{equation}
where $\mathcal{H}$ denotes the Hadamard operation. This means that our block encoding gives an extra factor of $1/\sqrt{2[(1-f(s))^2+f(s)^2]}$.
This prefactor is between $1/\sqrt{2}$ and 1, and will reduce the gap. %Thus the overall gap is reduced by a maximum factor of 2.

Then the complete circuit diagram can be given as in \cref{fig:block}. The walk operator $W_T(s)$ can be completed by a reflection about zero on all ancilla registers except $a_h$ which is part of the target system for the Hamiltonian.

\begin{figure}[H]
\centering
\includegraphics[width = 0.5\textwidth]{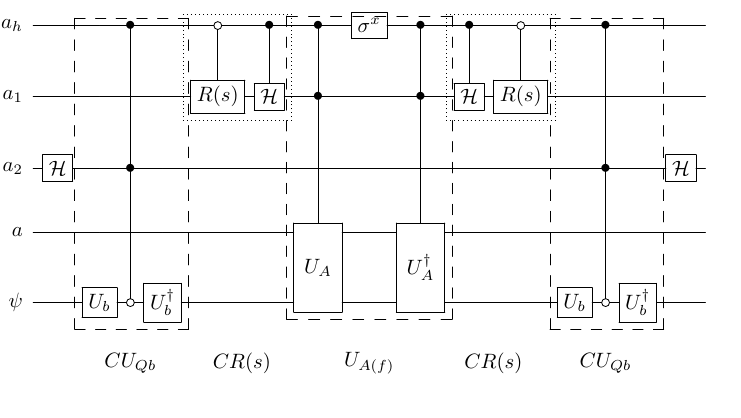}
\caption{The block encoding of the Hamiltonian $H(s)$, where the target system is labeled as $\ket\psi$, and $a_h$ as the extra qubit used to construct the Hamiltonian for the QLSP. The ancillas resulting from the block-encoded construction of $H(s)$ are labeled by $a_1$, $a_2$, and $a$ (where $a$ is the ancilla needed in the block encoding of $A$). The dashed boxes show $CU_{Qb}$ and $U_{A(f)}$, whereas the dotted boxes show $CR(s)$).
\label{fig:block}}
\end{figure}

\subsection{Non-Hermitian matrix $A$}

For the general case, we briefly review what is given in \cite{CostaAnYuvalEtAl2022} for the ancilla qubits needed, since we enlarge the space to address the non-Hermitian property of $A$. 

We have the ancilla qubit $a$ for the block encoding of $A$, as given for the PD case, and the remainder of the ancilla qubits are used as follows.
\begin{enumerate}
    \item The first selects between the blocks in $A(f)$.
    \item The next is used for preparing the combination of 
    $\sigma_z \otimes I$ and $A$.
    \item The third is used in implementing $Q_b$.
    \item The fourth selects between the blocks in $H(s)$.
\end{enumerate}
The first and fourth ancillas above are part of the target system for the Hamiltonian $H(s)$, and are denoted $a_{h_1}$ and $a_{h_2}$, respectively.
The second and third qubits in the list above are denoted $a_1$ and $a_2$, respectively.
%We named the ancillas that inherit from the Hamiltonian given in \cref{eq:Hsencoding} by $a_{h_1}$, the first qubit listed above, and $a_{h_2}$ for the last one. For the second and the third qubits, we named by $a_1$ and $a_2$, respectively.
The circuit diagram for the block encoding for the general case is presented in \cref{fig:block2}.
See Ref.~\cite{CostaAnYuvalEtAl2022} for its complete derivation.

\begin{figure}[H]
\centering
\includegraphics[width = 0.5\textwidth]{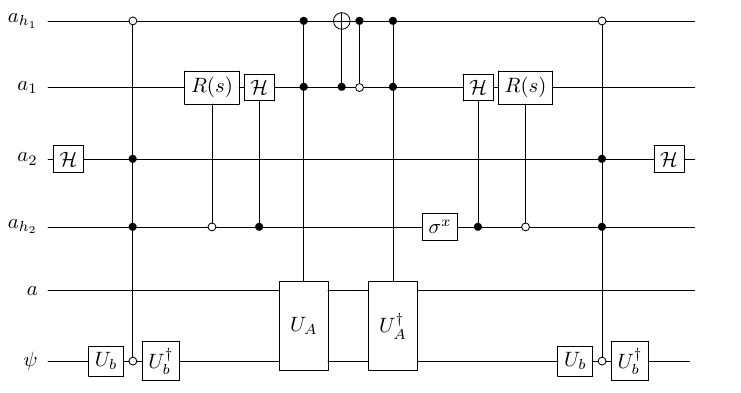}
\caption{The block encoding of the Hamiltonian $H(s)$ for the non-Hermitian case, where again the target system is labeled as $\ket\psi$, and the ancilla are labeled by $a_{h_1}$, $a_1$, $a_2$,  $a_{h_2}$ and $a$ (where $a_{h_1}$ and $a_{h_2}$ are part of the system acted upon by the Hamiltonian).
\label{fig:block2}}
\end{figure}

\section{Numerical Tests - Supporting Material}
\label{app:support}

\subsection{QW method}

In Fig.~\ref{fig:AQC_PD} (positive definite) and Fig.~\ref{fig:AQC_Nherm} (non-Hermitian)  we show the results from \cref{fig:comparison_geo_mean} for dimension $16\times 16$ matrices, as well as results for $4\times 4$ and $8\times 8$ matrices. 
In these figures, we show the results for the three dimensions in subfigures as well and find that
the dimension does not significantly alter the complexity of the QLSP, as we expect from theory. 

The table for the PD for a fixed total number of steps for all the 100 instances tested for each condition number is included in \cref{tab:Avg}.
The worst relative performance of the QW method compared to the RM is for matrices with condition number equal to 10, but the QW performance is still 13 times better. 

Moreover, we observed a large amount of outlier instances in the quantum walk for the non-Hermitian matrices where the solution is obtained after a small number of steps. This occurs due to the relatively large allowed error $\Delta = 0.4$ because when taking sufficiently large number of samples, one can be sufficiently close to
the solution purely by chance.  Such outliers are expected when sampling randomly, especially at low step counts, and do not contradict the broader trend that the discrete adiabatic solver consistently requires fewer resources across all instances.

\begin{table}[tbh]
\centering
\begin{center}
\begin{tabular}{ |c|c|c| } 
 \hline
 Condition Number & Number of steps  & $\Delta$ \\ 
 \hline
$10$ & $4$ &  $0.378$
\\ 
$20$ & $12$ & $0.310$ 
\\ 
$30$ & $16$ &$0.339$ 
\\ 
$40$ & $20$ &$0.353$
\\ 
$50$ & $24$ & $0.362$ 
\\
\hline
\end{tabular}
\end{center}
\caption{Number of steps for average error $\Delta$ $\sim 0.4$ with a range of condition numbers for PD matrices of dimension $16\times 16$. The RMS of the error (as measured by the norm of the difference in states), is computed over 100 instances for each condition number considered.}
\label{tab:Avg}
\end{table}

\begin{figure}[tbh]
    \centering
    \begin{tabular}{@{}p{0.8\linewidth}}
    \subfigimg[width=\linewidth]{(a)}{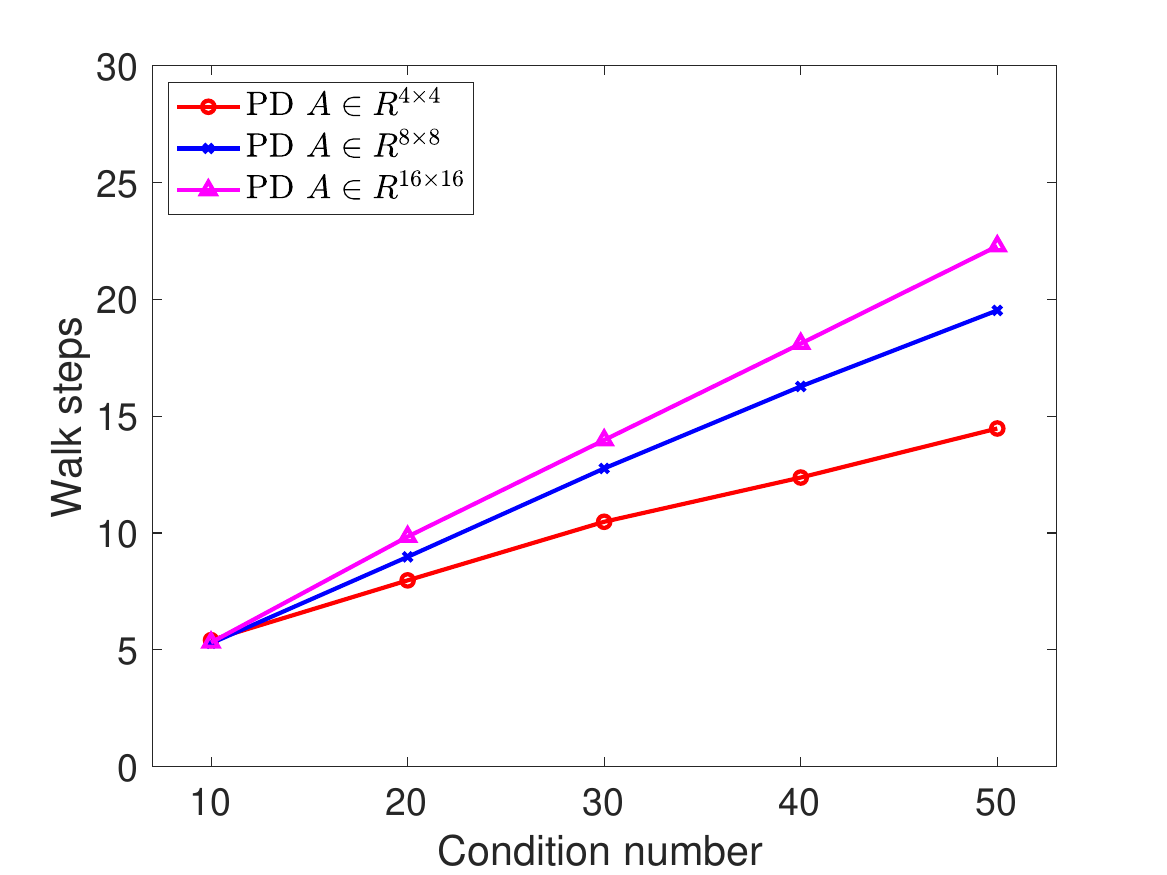} \\
    \subfigimg[width=\linewidth]{(b)}{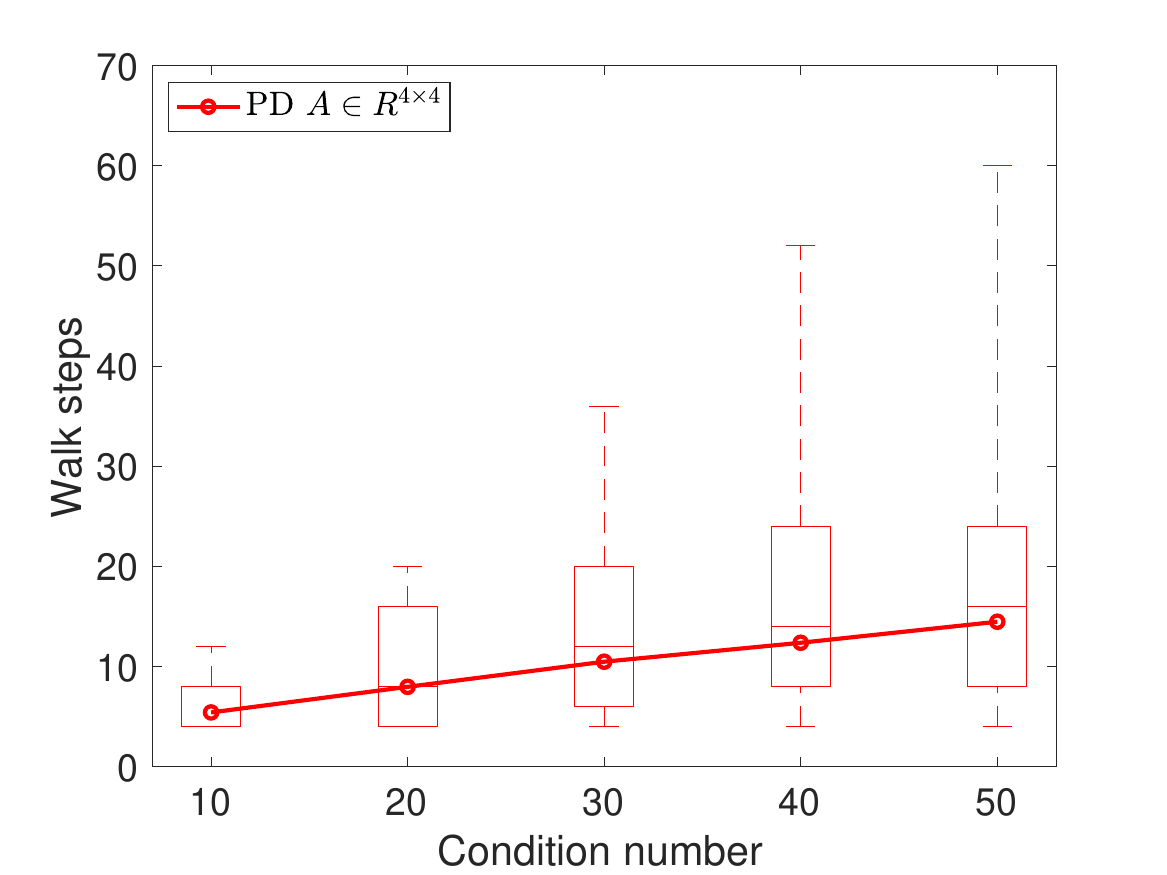} \\
    \subfigimg[width=\linewidth]{(c)}{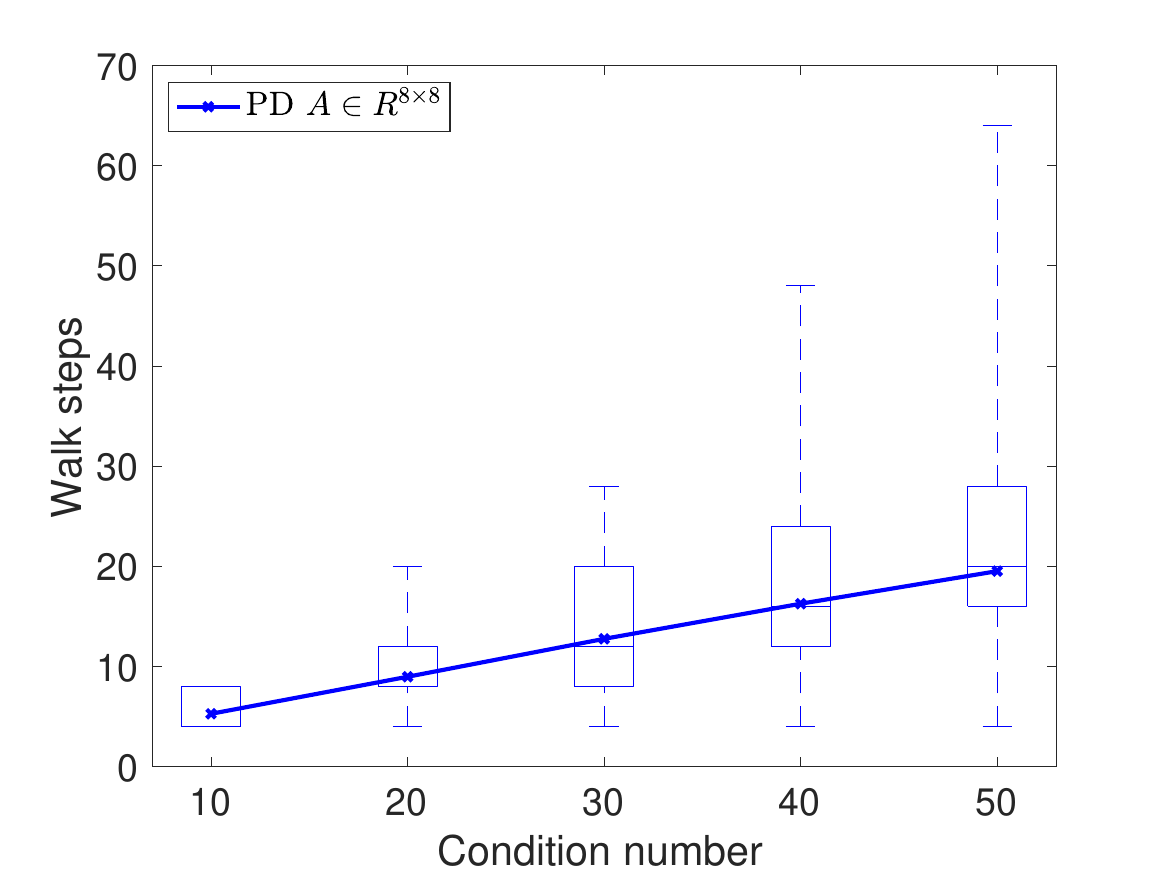} \\
    \subfigimg[width=\linewidth]{(d)}{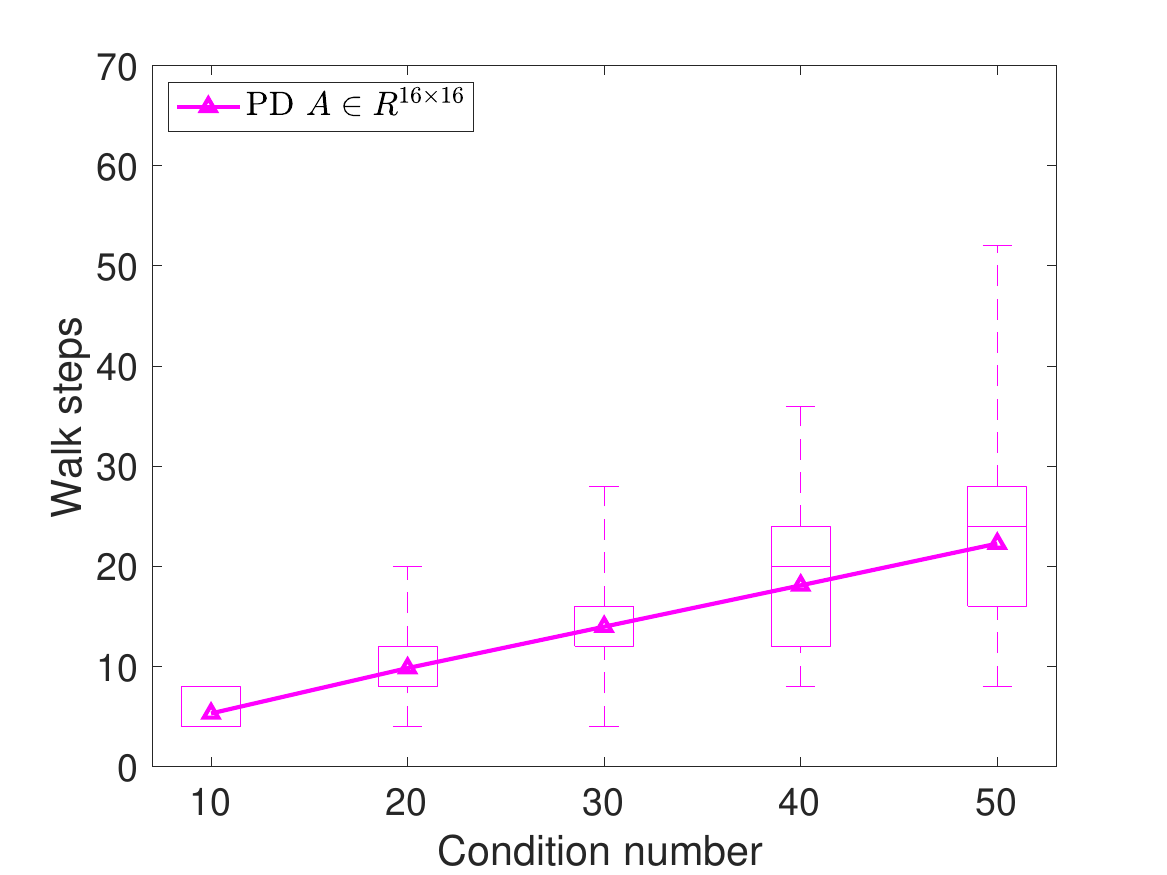} \\
    \end{tabular}
    \caption{Numerical tests of the QW method with positive definite matrices of dimension $4\times 4$, $8\times 8$, and $16\times 16$. The vertical axis shows the required total walk steps with $\Delta=0.4$ as the threshold error. Part (a) shows the averaged results of $100$ randomly generated instances, and parts (b) to (d) show the box plots.}
    \label{fig:AQC_PD}
\end{figure}

\begin{figure}[tbh]
    \centering
    \begin{tabular}{@{}p{0.8\linewidth}}
    \subfigimg[width=\linewidth]{(a)}{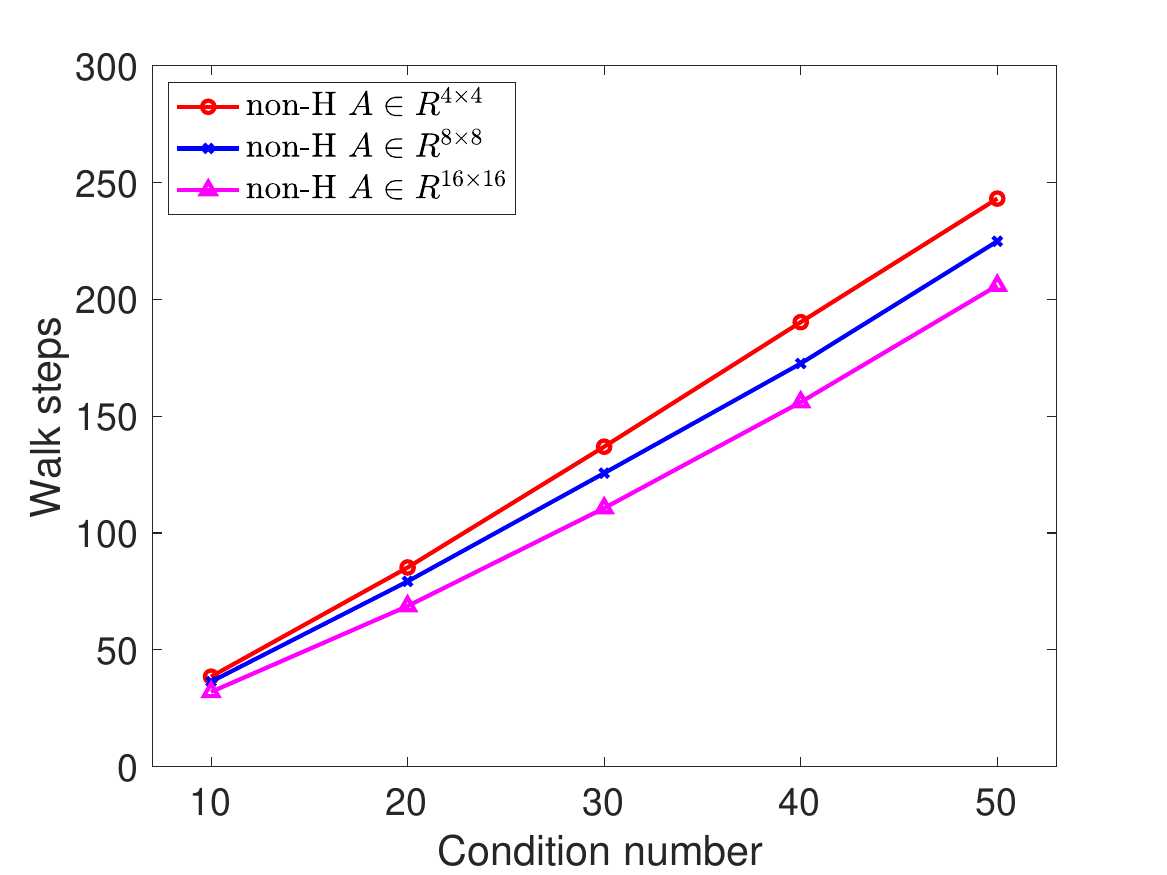} \\
    \subfigimg[width=\linewidth]{(b)}{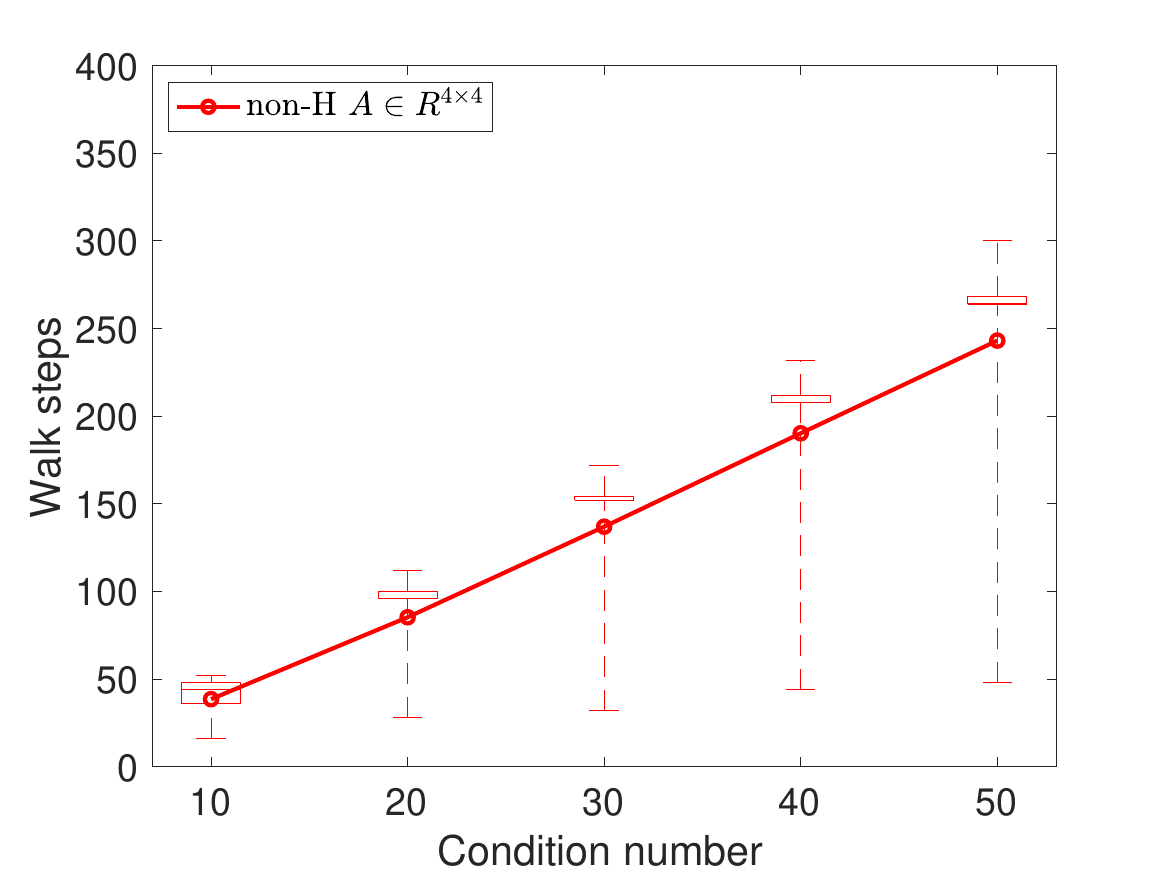} \\
    \subfigimg[width=\linewidth]{(c)}{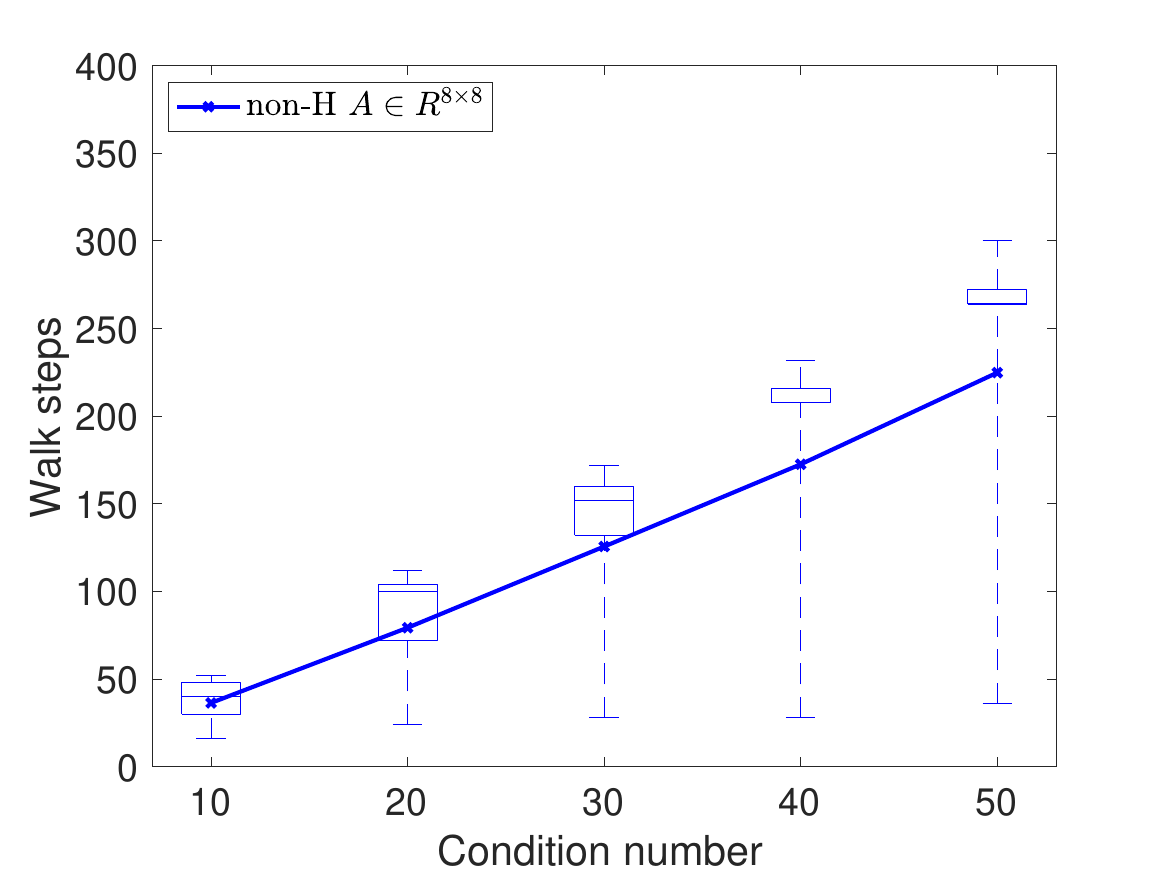} \\
    \subfigimg[width=\linewidth]{(d)}{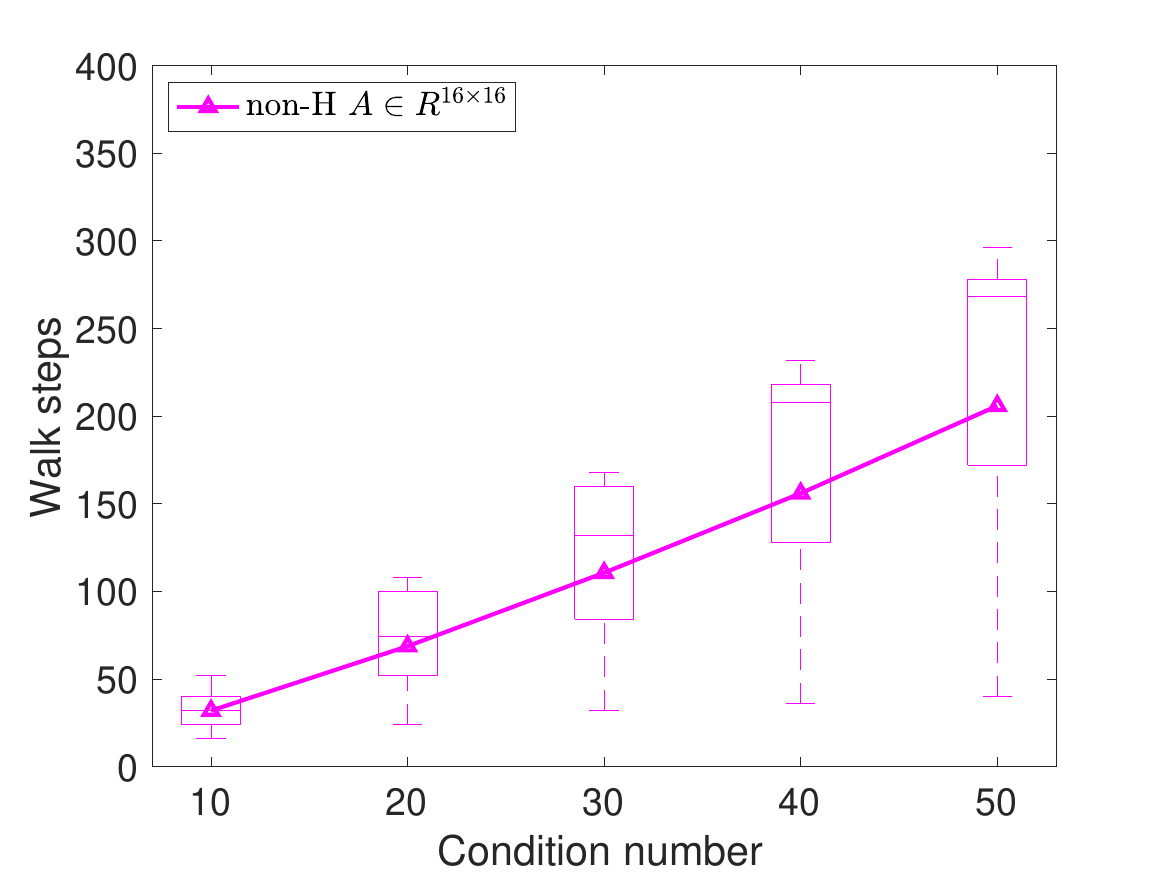} \\
    \end{tabular}
    \caption{Numerical tests of the QW method with non-Hermitian matrices of dimension $4\times 4$, $8\times 8$, and $16\times 16$. The vertical axis shows the required total walk steps with $\Delta=0.4$ as the threshold error. 
    Part (a) shows the averaged results of $100$ randomly generated instances, and parts (b) to (d) show the box plots.}
    \label{fig:AQC_Nherm}
\end{figure}

\subsection{Randomized Method}

In Fig.~\ref{fig:RM_PD} (positive definite) and Fig.~\ref{fig:RM_nonH} (non-Hermitian) we show results from \cref{fig:comparison_geo_mean} for dimension $16\times 16$, and also dimension $4\times 4$ and $8\times 8$ matrices.
There is even less dependence of the complexity on the dimension than for the QW method.
The dimension dependence in the complexity comes from the matrix encoding, which does not influence the numerical analysis.

We also include the table (\cref{tab:Avg_RM}) that gives the RMS for the target error for the PD matrices, when for each condition number we fix the same averaged time for all instances of matrices tested.

\begin{table}[tbh]
\centering
\begin{center}
\begin{tabular}{ |c|c|c| } 
 \hline
 Condition Number & Averaged time  & $\Delta$\\ 
 \hline
$10$ & $37.2 $ &  $0.379$
\\ 
$20$ & $88.1$  & $0.395$
\\ 
$30$ & $144.5$ & $0.396 $ 
\\ 
$40$ & $202.7$ & $0.398$ 
\\ 
$50$ & $270.9$ & $0.392$ 
\\
\hline
\end{tabular}
\end{center}
\caption{Average evolution time for average error $\Delta$ $\sim 0.4$ via RM with a range of condition numbers for PD matrices of dimension $16\times 16$.
The RMS error is computed over 100 instances for each condition number considered and 200 repetitions per instance.}
\label{tab:Avg_RM}
\end{table}

\begin{figure}[tbh]
    \centering
     \begin{tabular}{@{}p{0.8\linewidth}}
    \subfigimg[width=\linewidth]{(a)}{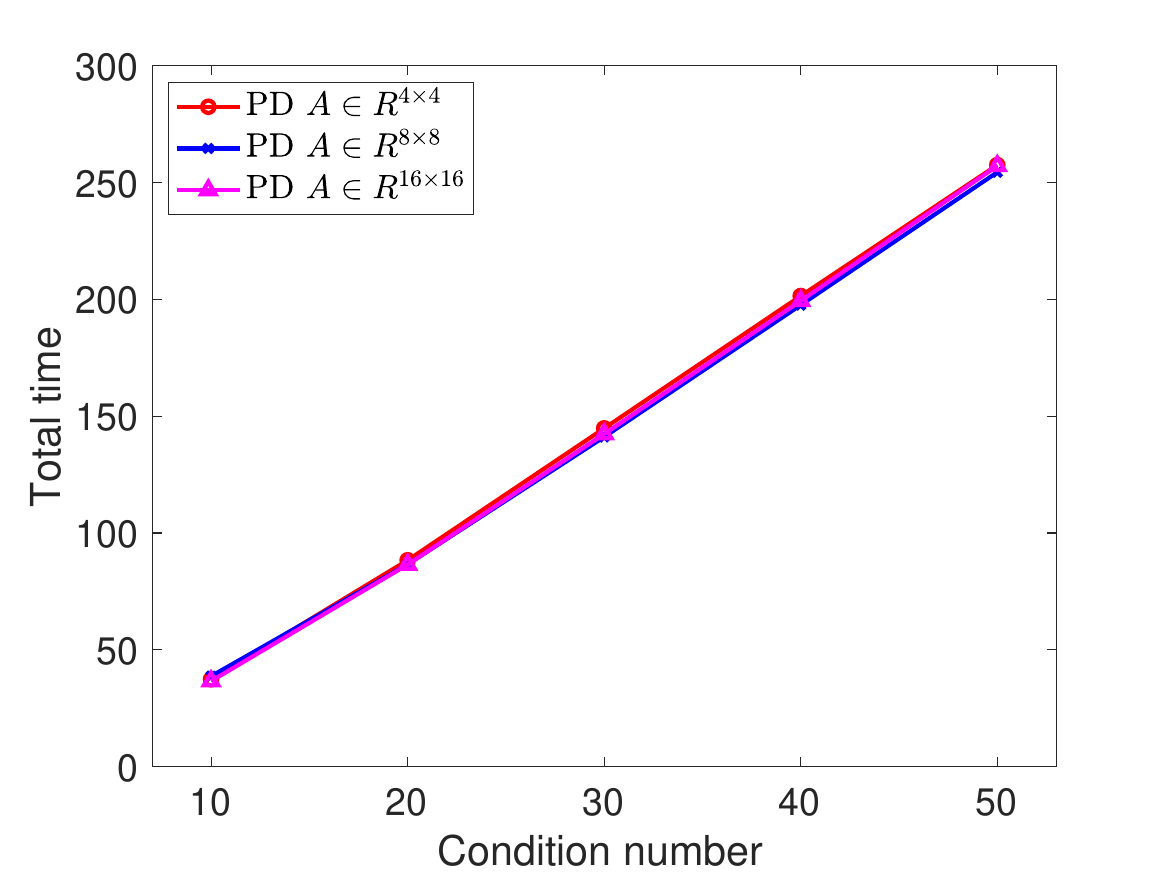} \\
    \subfigimg[width=\linewidth]{(b)}{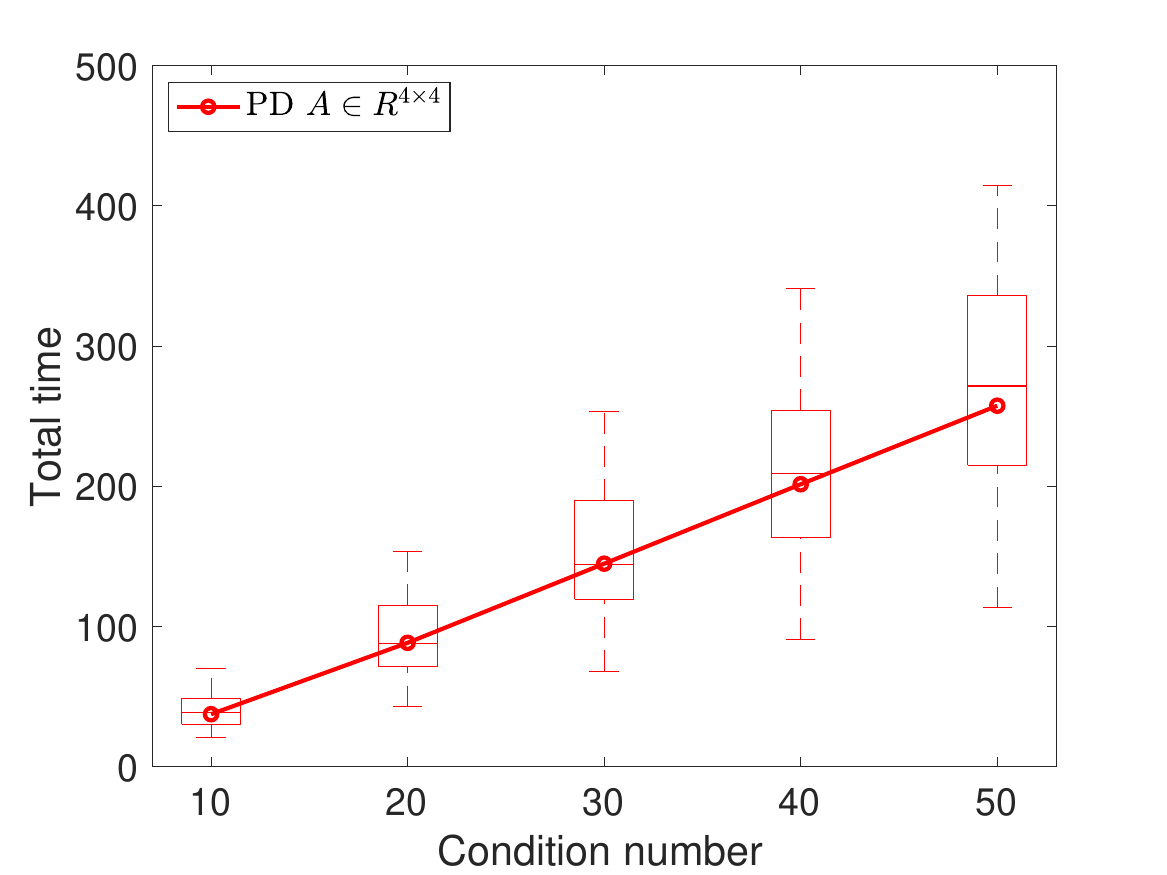} \\
    \subfigimg[width=\linewidth]{(c)}{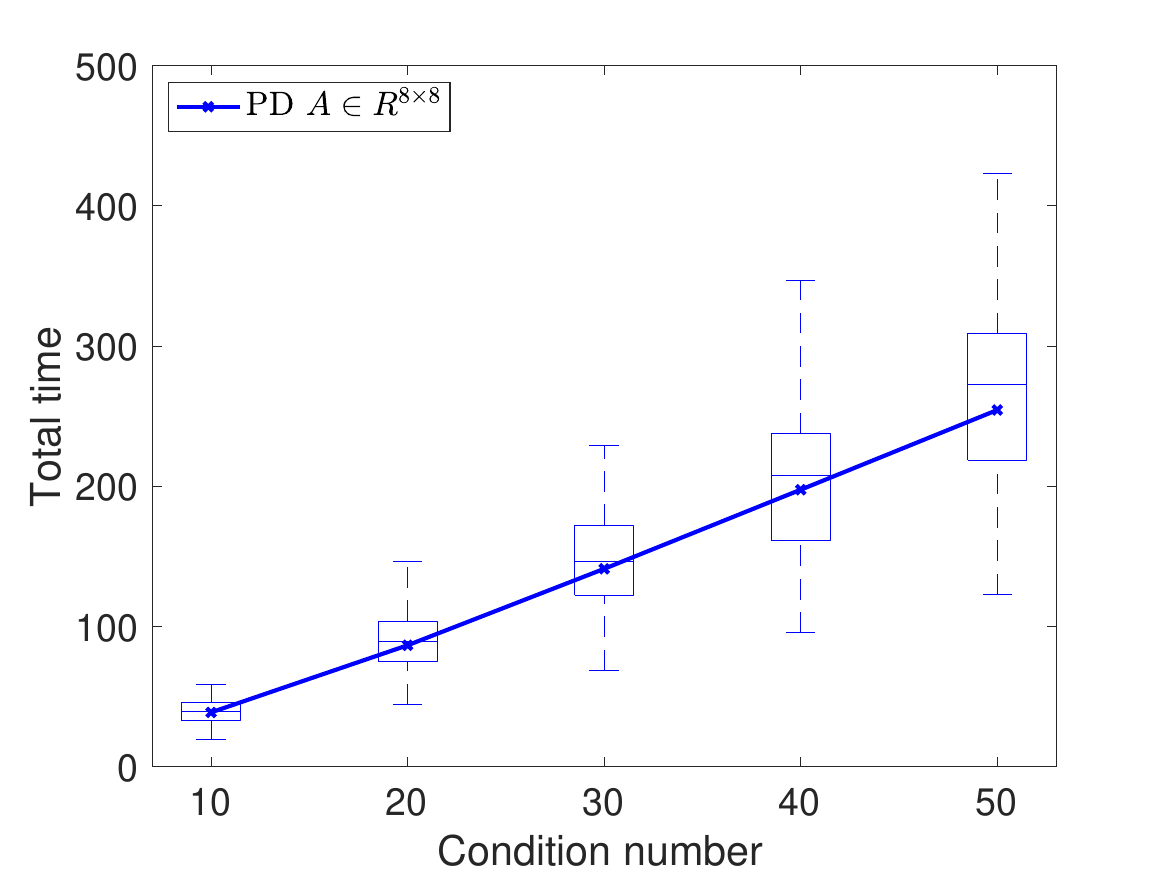} \\
    \subfigimg[width=\linewidth]{(d)}{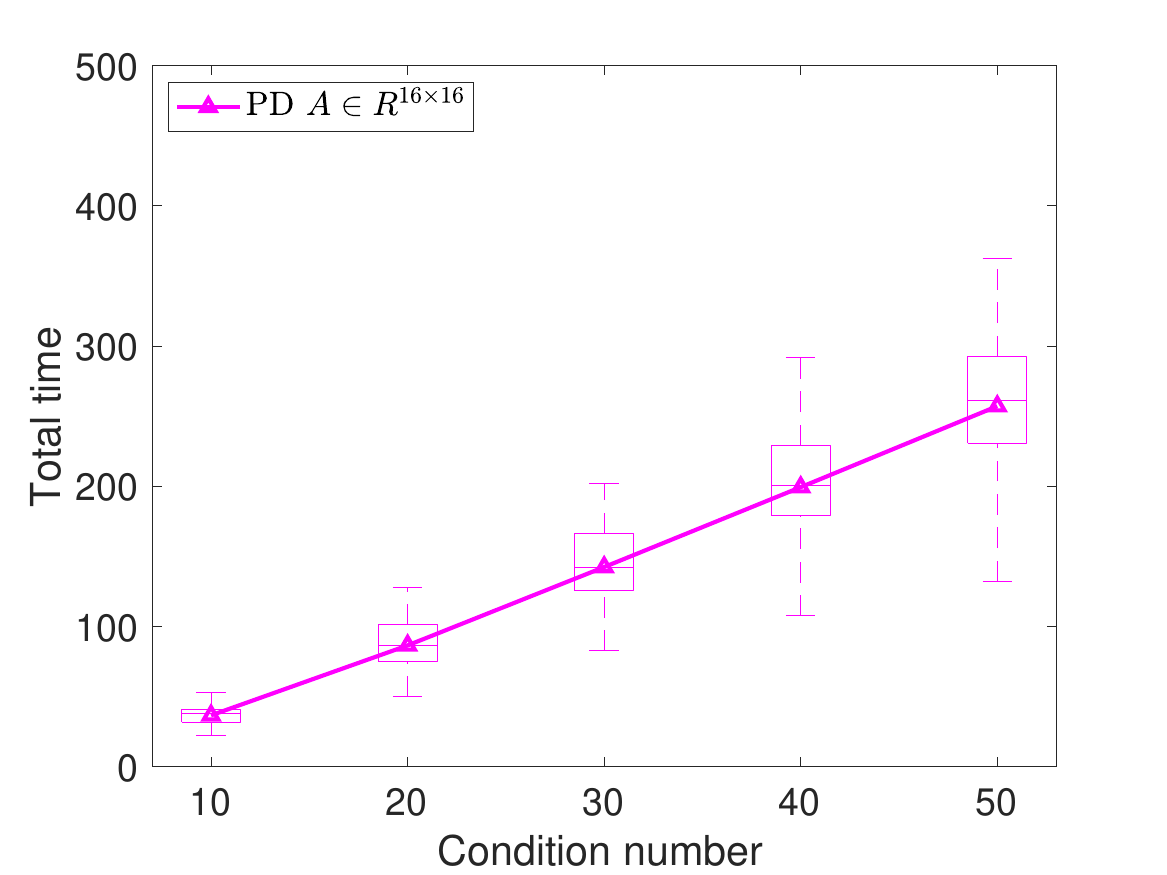} \\
    \end{tabular}
    \caption{Numerical tests of the Randomized method with positive definite matrices of dimension $4\times 4$, $8\times 8$, and $16\times 16$. The vertical axis shows the required total time (defined as $\sum_{j=1}^q |t_j|$) to reduce the RMS below $0.4$. 
    Part (a) shows the averaged results of $100$ randomly generated instances, and parts (b) to (d) show the box plots.}
    \label{fig:RM_PD}
\end{figure}

\begin{figure}[tbh]
    \centering
    \begin{tabular}{@{}p{0.8\linewidth}}
    \subfigimg[width=\linewidth]{(a)}{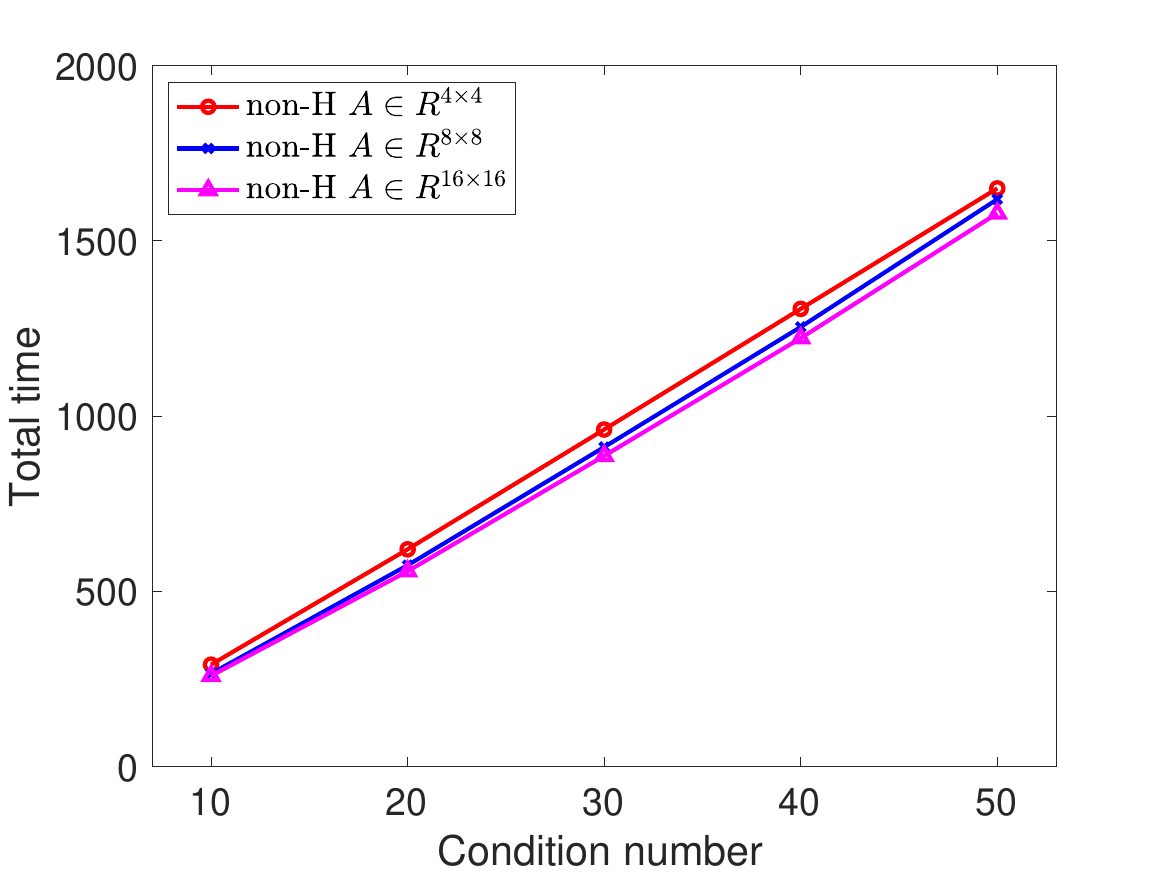} \\
    \subfigimg[width=\linewidth]{(b)}{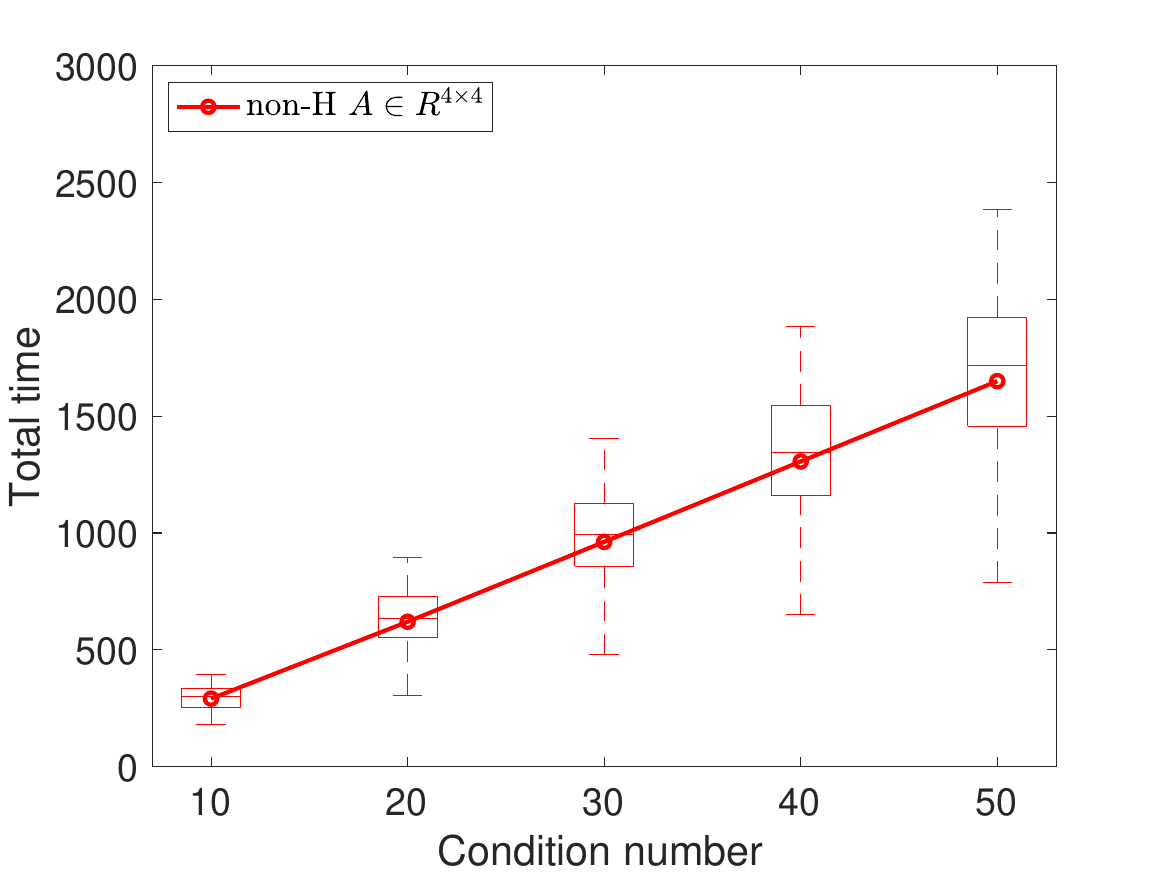} \\
    \subfigimg[width=\linewidth]{(c)}{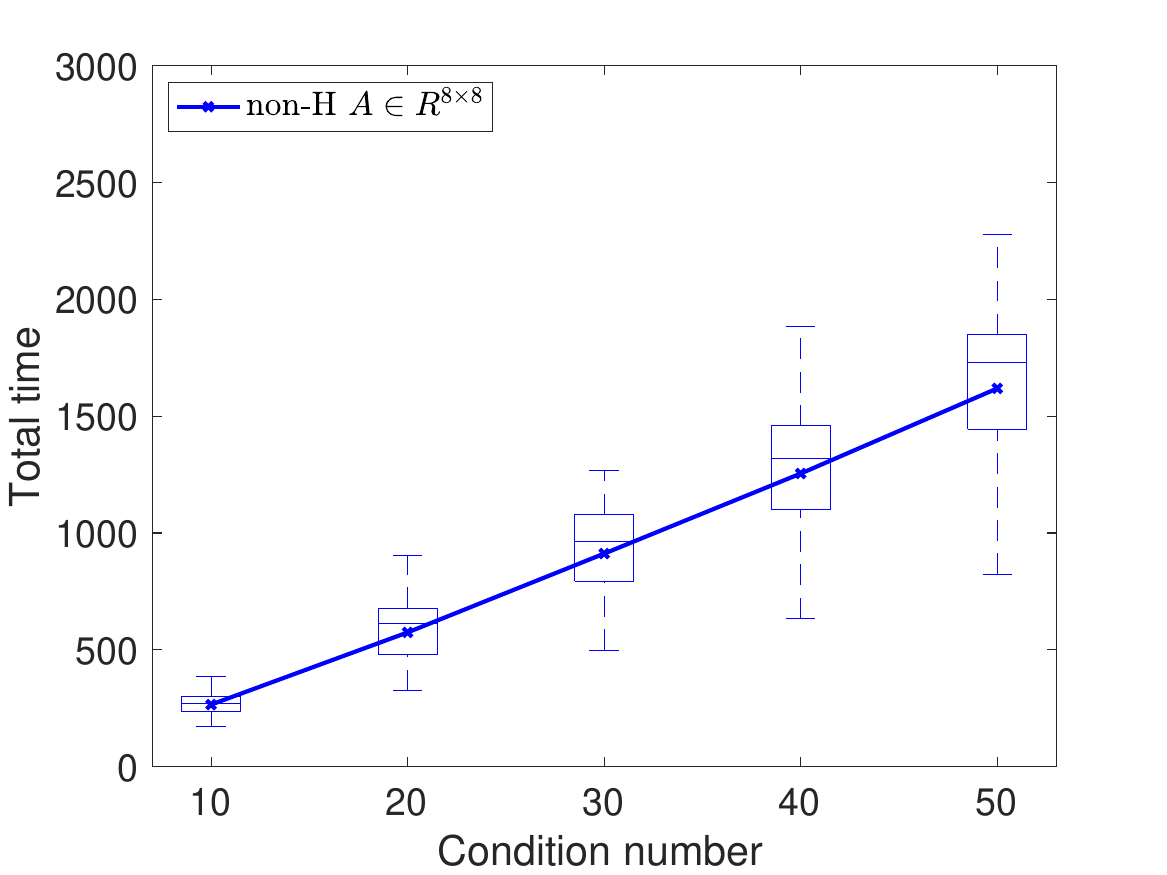} \\
    \subfigimg[width=\linewidth]{(d)}{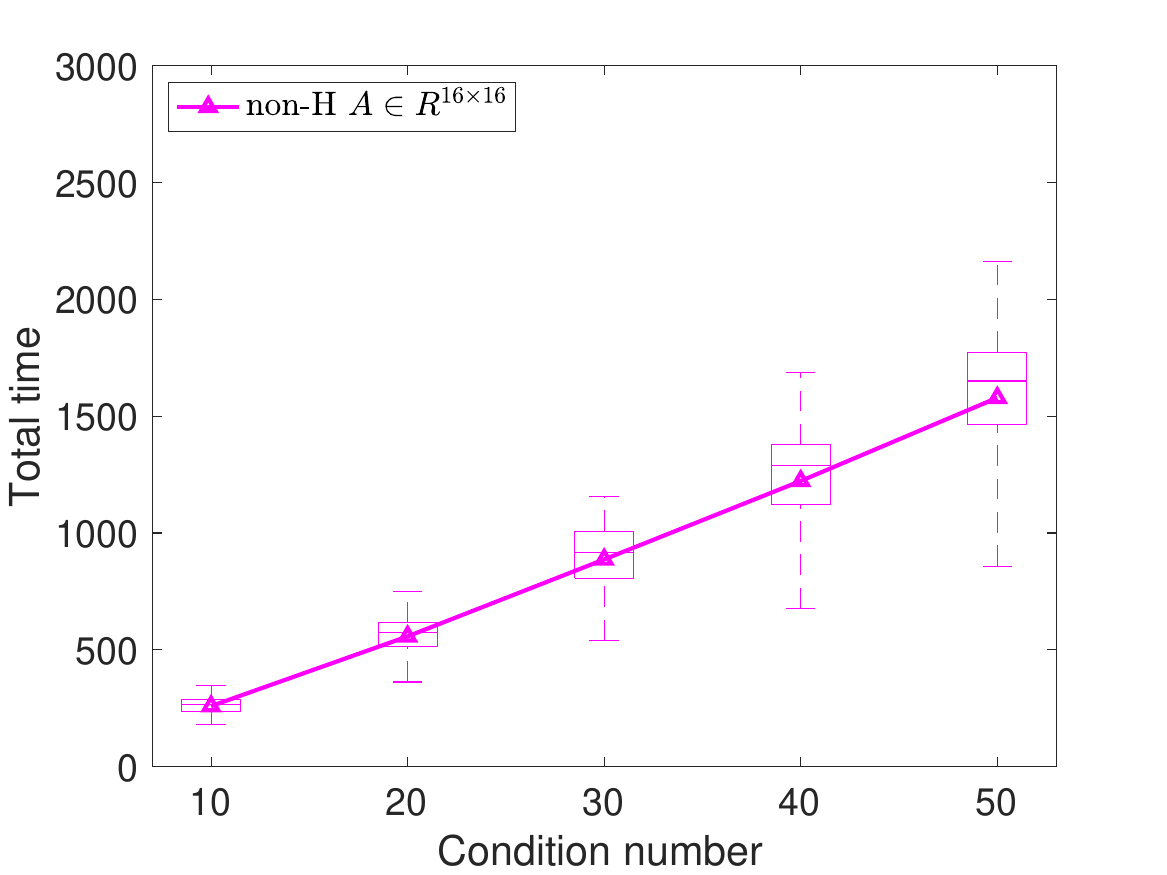} \\
    \end{tabular}
    \caption{Numerical tests of the Randomized method with non-Hermitian matrices of dimension $4\times 4$, $8\times 8$, and $16\times 16$. The vertical axis shows the required total time (defined as $\sum_{j=1}^q |t_j|$) to reduce the RMS below $0.4$.
    Part (a) shows the averaged results of $100$ randomly generated instances, and parts (b) to (d) show the box plots.}
    \label{fig:RM_nonH}
\end{figure}

\subsection{QW vs RM - scatter plots}
In this section, we compare the performance between the RM and the QW method of each instance tested for $A\in \mathbb{R}^{8\times 8}$ for PD and Hermitian matrices in \cref{fig:scattering} and non-Hermitian matrices in \cref{fig:scattering2}. The vertical axis shows the number of walk steps of the instances to obtain the desired solution error $\Delta=0.4$. In the horizontal axis, we have the average minimal time of the Randomized method to reduce the RMS error below $\Delta=0.4$. 

In the numerical results, there were no tested cases where the Randomized method surpassed the QW performance.
This can be seen from the fact that the complexity for all instances is below the dashed line in \cref{fig:scattering} and \cref{fig:scattering2}.
We even find that the QW still wins if we compare different problem instances with the same condition number.
That is, the maximum cost of the QW is below the minimum cost of the RM (provided we select matrices of the same type, either both PD or both non-Hermitian).

The same can be concluded from the cases of other dimensions by looking at the error bars, since they do not overlap.

\begin{figure}[tbh]
    \centering
    \includegraphics[width = 0.32\textwidth]{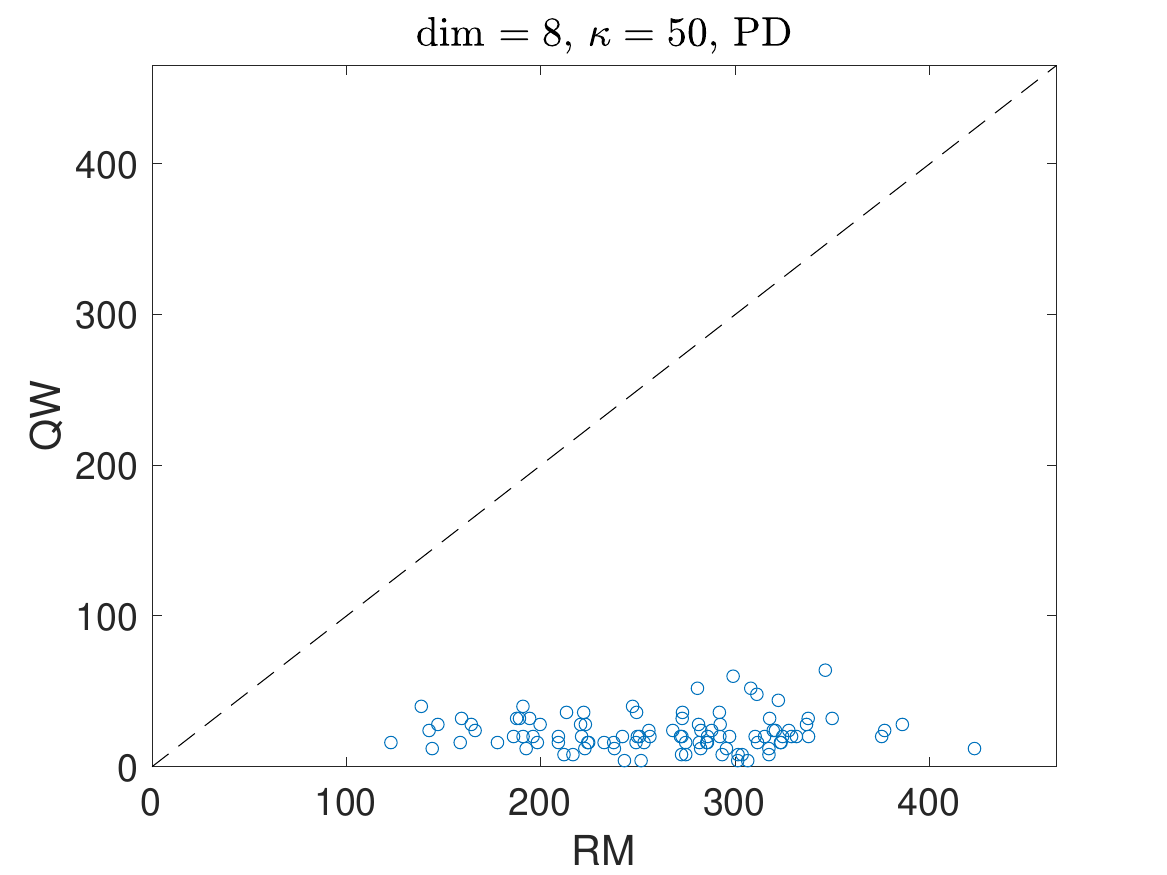}\\
    \includegraphics[width = 0.32\textwidth]{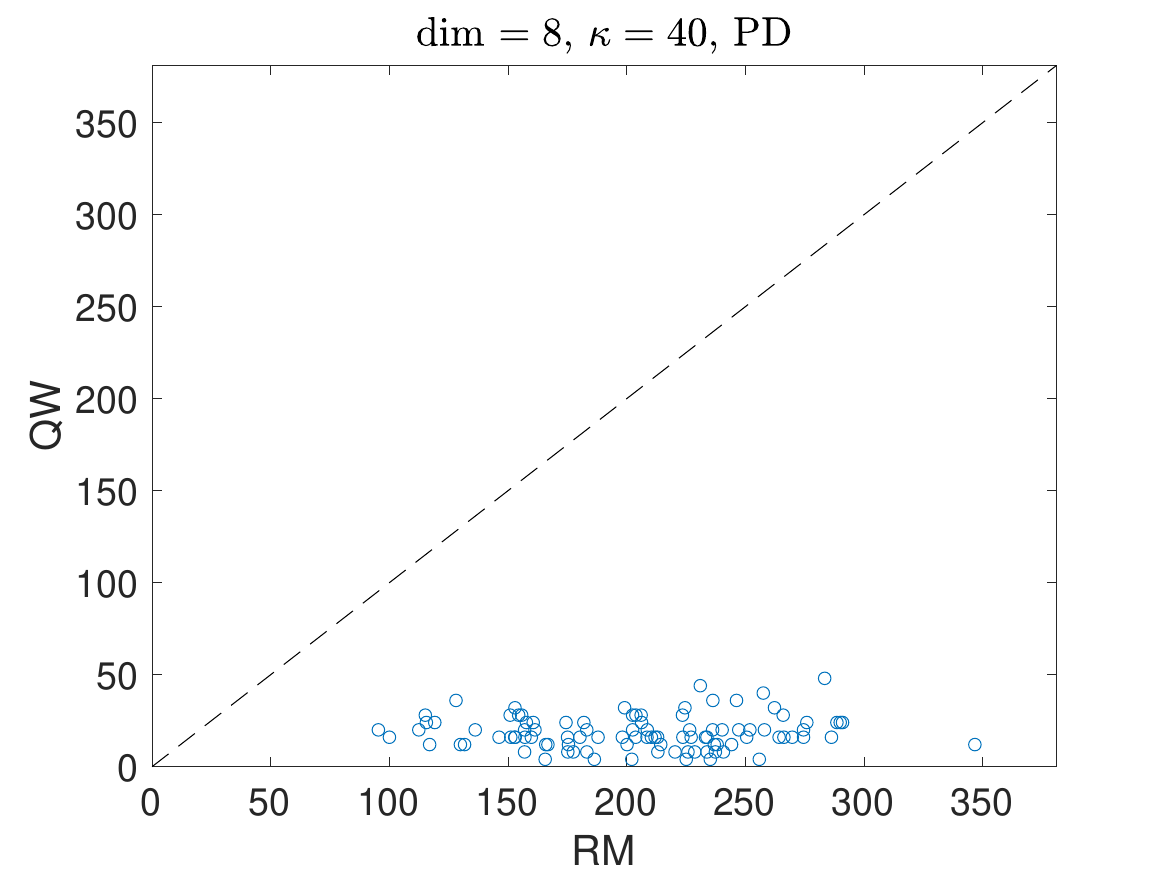}\\
    \includegraphics[width = 0.32\textwidth]{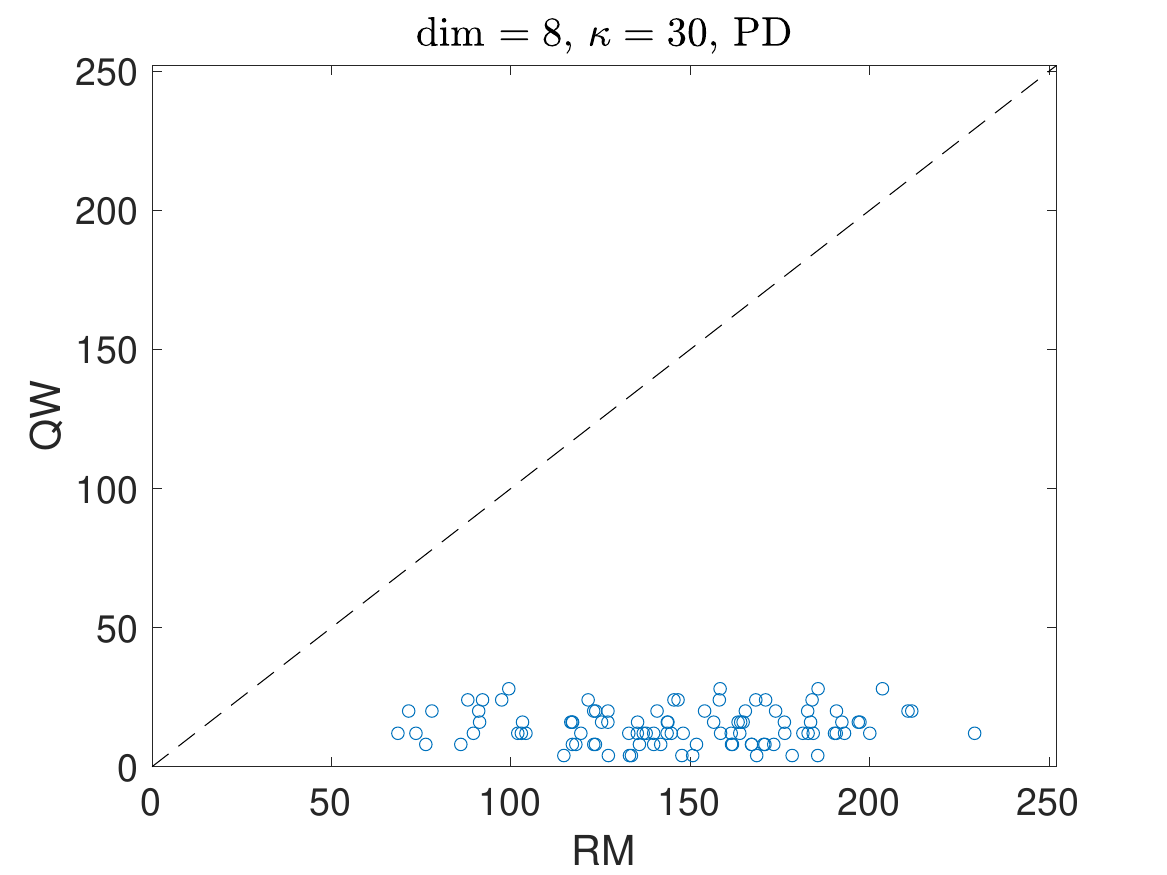}\\
    \includegraphics[width = 0.32\textwidth]{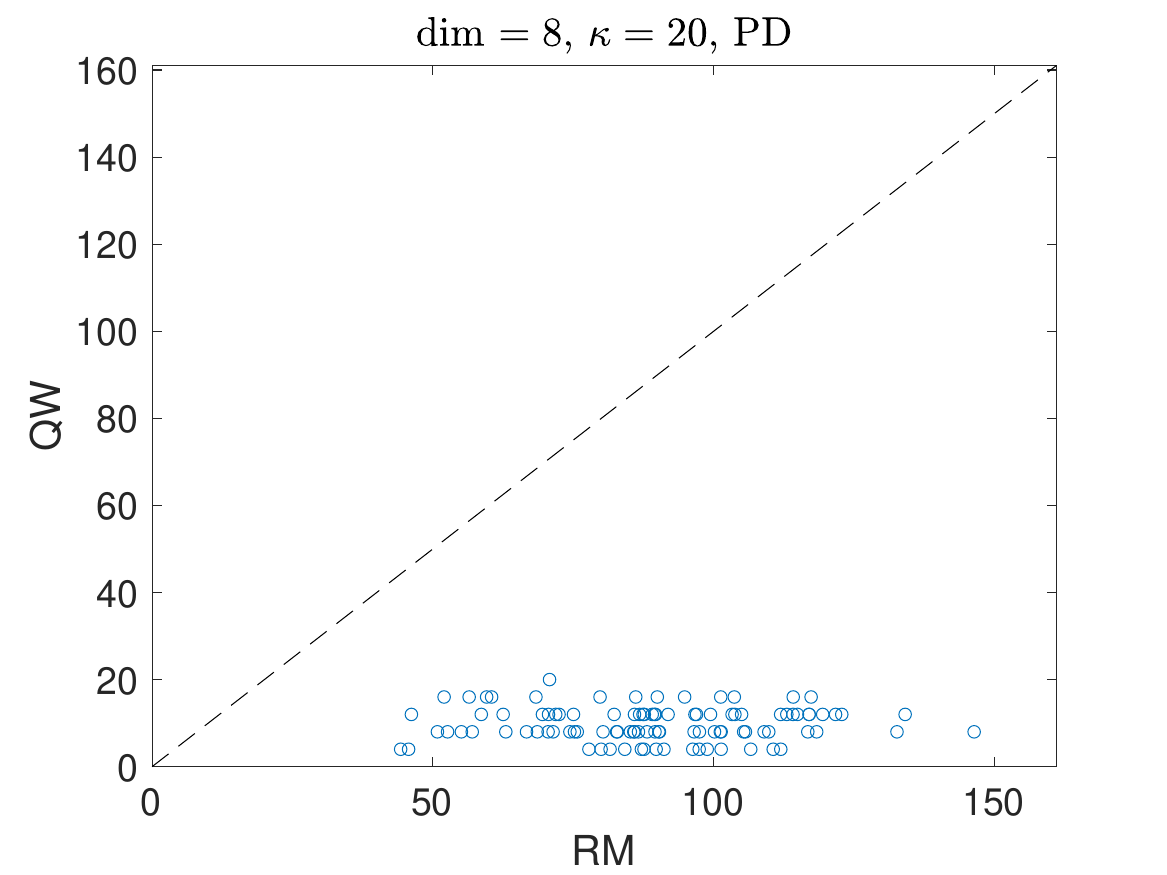}\\
    \includegraphics[width = 0.32\textwidth]{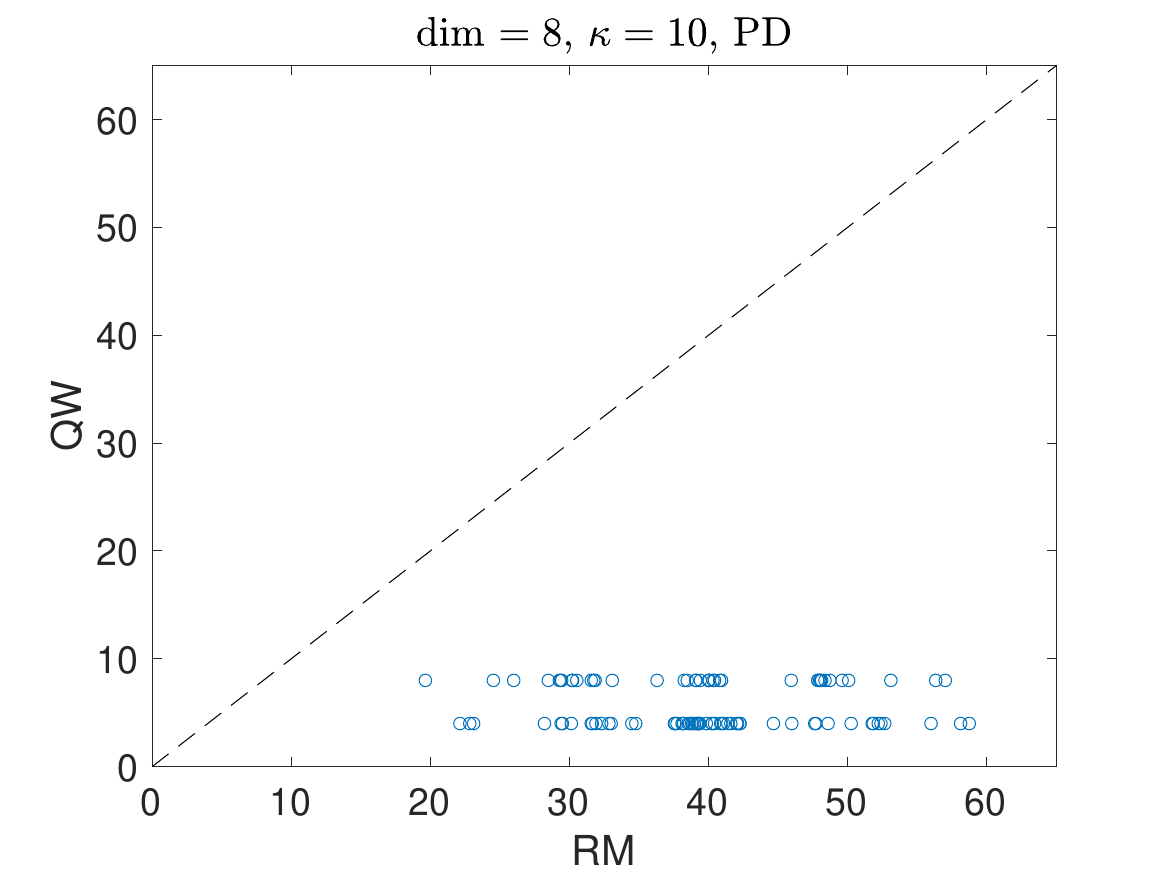} 
    \caption{Scatter plots for the positive definite instances used in testing QW and RM. In the vertical axis, we have the complexity (number of walk steps) for the quantum walk method, and in the horizontal axis the complexity (evolution time) for the Randomized method.}
    \label{fig:scattering}
\end{figure}

\begin{figure}[tbh]
    \centering
    \includegraphics[width = 0.32\textwidth]{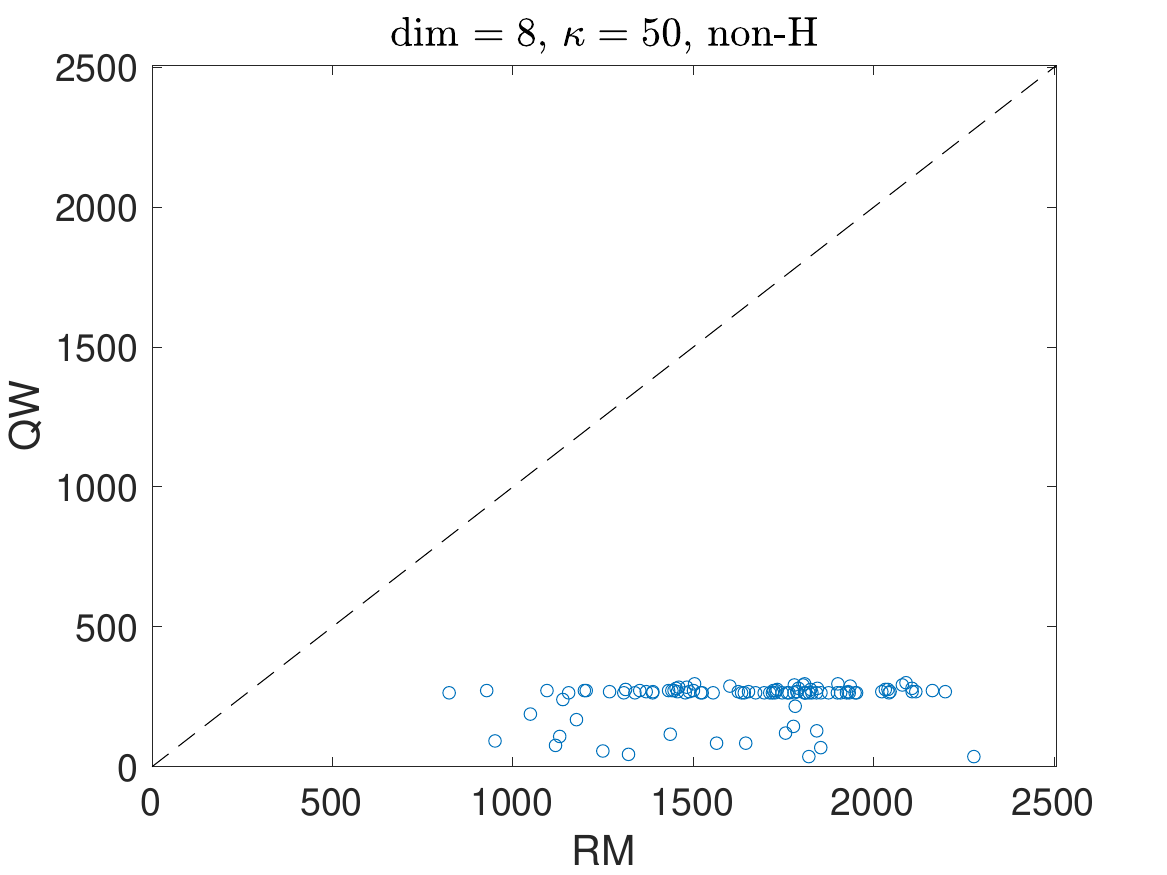}\\
    \includegraphics[width = 0.32\textwidth]{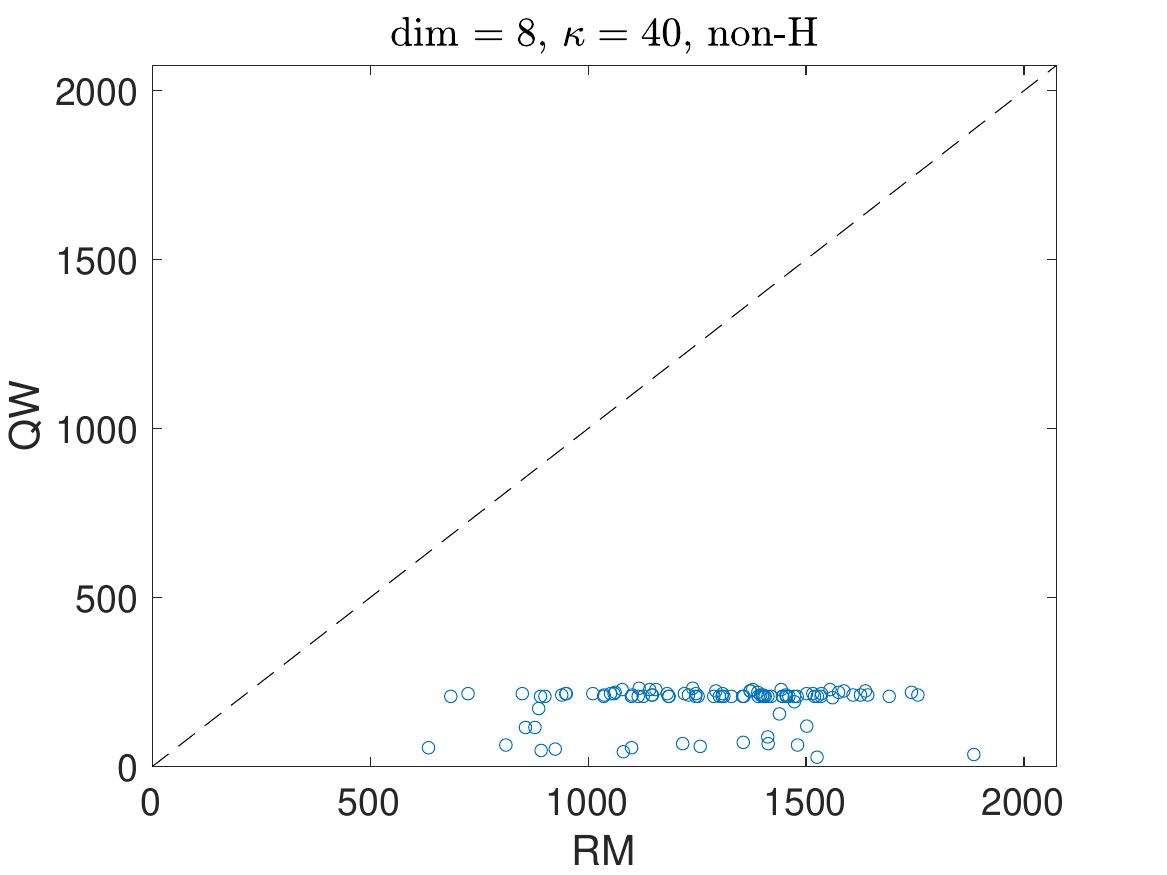}\\
    \includegraphics[width = 0.32\textwidth]{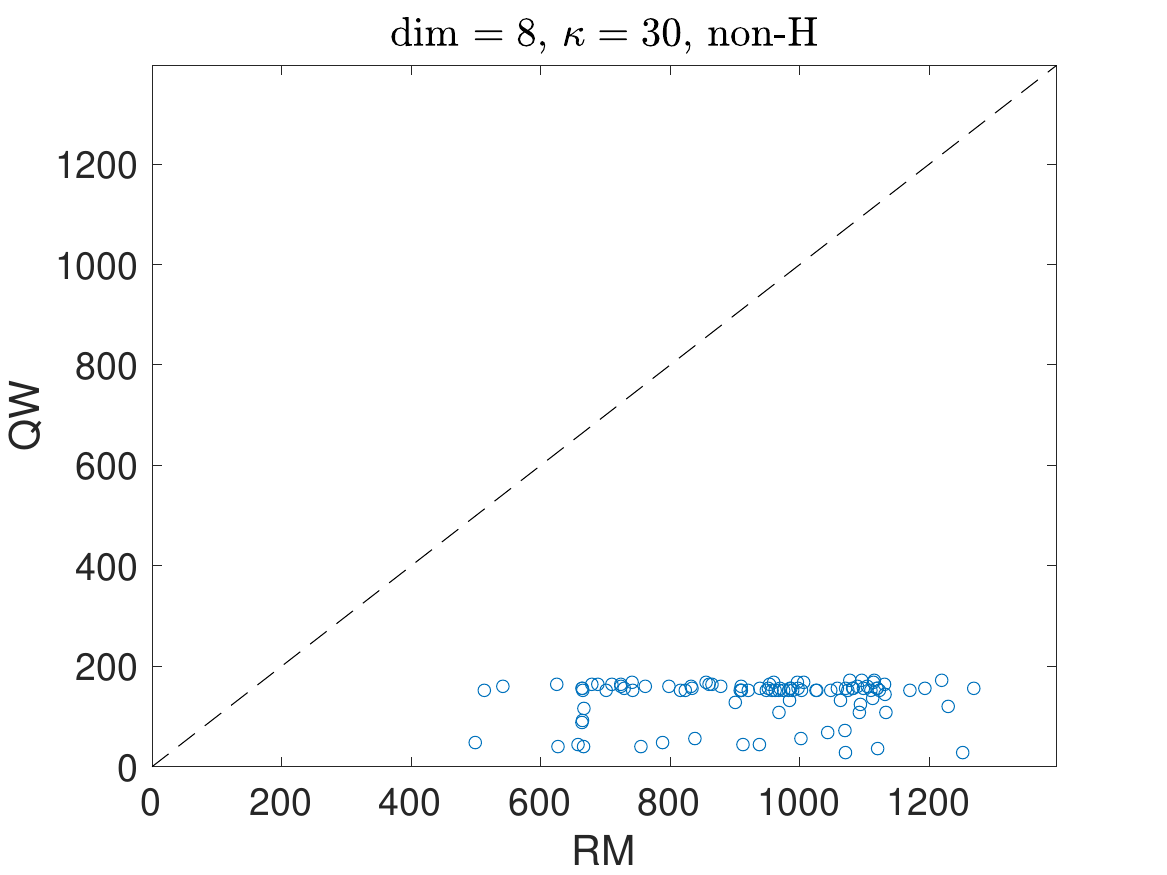}\\
    \includegraphics[width = 0.32\textwidth]{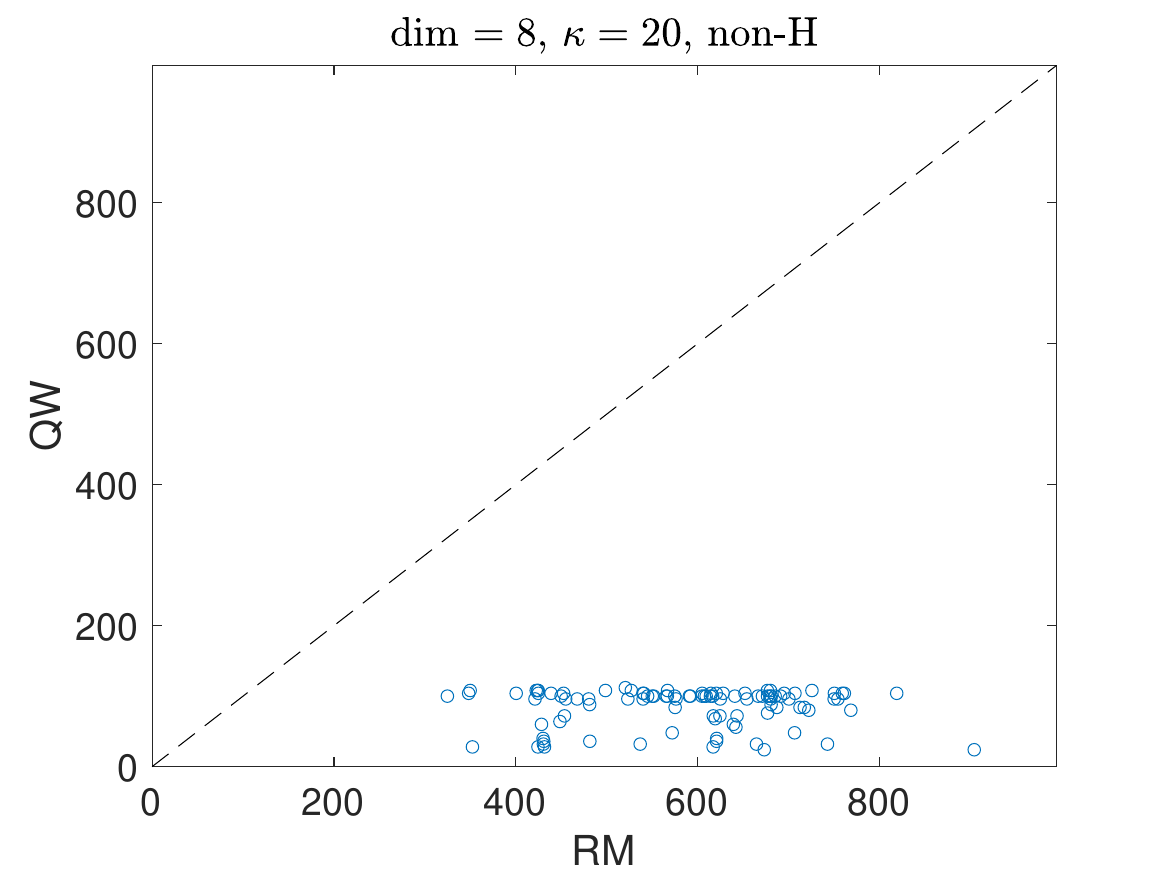}\\
    \includegraphics[width = 0.32\textwidth]{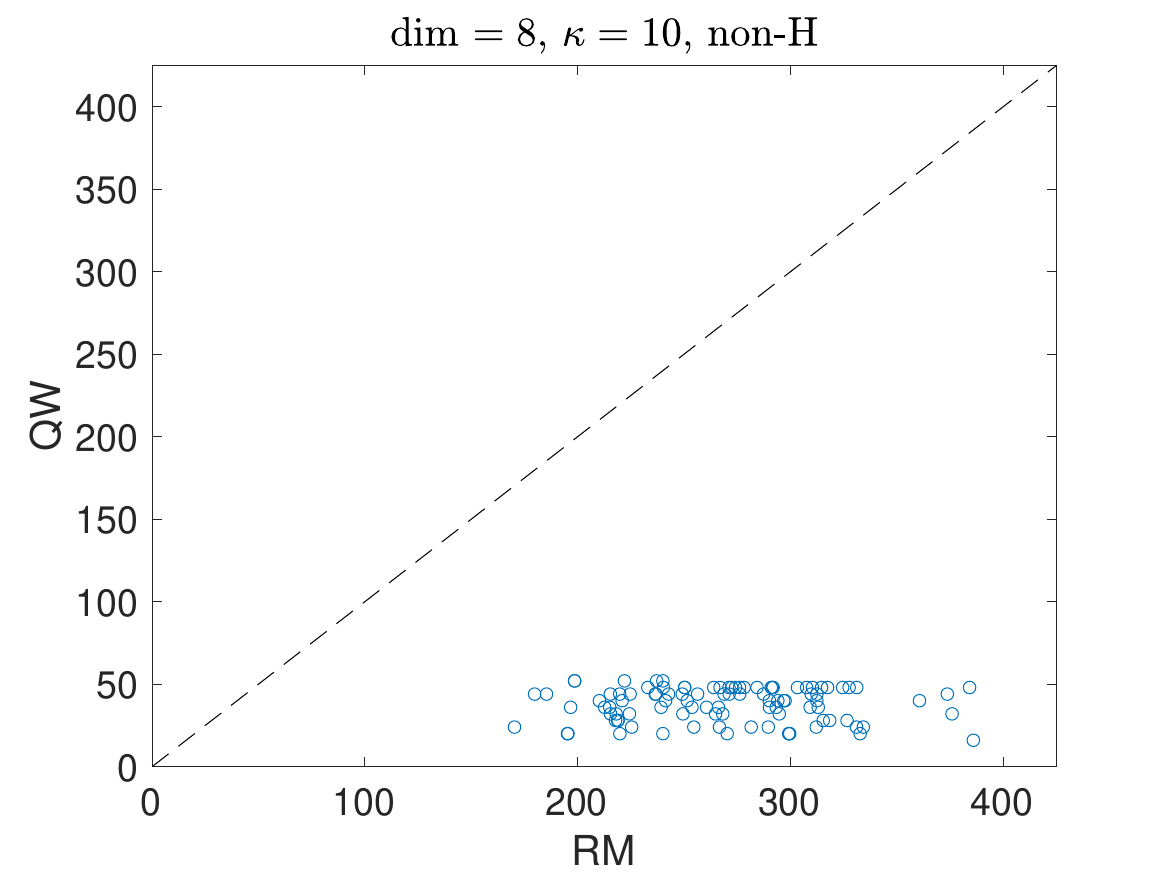}
    \caption{Scatter plots for the non-Hermitian instances used in testing QW and RM.
    The vertical axis is the complexity (number of walk steps) for the quantum walk method, and the horizontal axis is the complexity (evolution time) for the Randomized method.}
    \label{fig:scattering2}
\end{figure}

\clearpage


\begin{thebibliography}{10}

\bibitem{Harrow_2009}
Aram~W. Harrow, Avinatan Hassidim, and Seth Lloyd.
\newblock ``Quantum algorithm for linear systems of equations''.
\newblock \href{https://dx.doi.org/10.1103/physrevlett.103.150502}{Physical
  Review Letters {\bf 103}, 150502}~(2009).

\bibitem{an2022quantum}
Dong An and Lin Lin.
\newblock ``Quantum linear system solver based on time-optimal adiabatic
  quantum computing and quantum approximate optimization algorithm''.
\newblock \href{https://dx.doi.org/10.1145/3498331}{ACM Transactions on Quantum
  Computing {\bf 3}, 5}~(2022).

\bibitem{ambainis2010variable}
Andris Ambainis.
\newblock ``Variable time amplitude amplification and a faster quantum
  algorithm for solving systems of linear equations''~(2010).
\newblock
  url:~\href{https://arxiv.org/abs/1010.4458}{arxiv.org/abs/1010.4458}.

\bibitem{Lin2020optimalpolynomial}
Lin Lin and Yu~Tong.
\newblock ``Optimal polynomial based quantum eigenstate filtering with
  application to solving quantum linear systems''.
\newblock \href{https://dx.doi.org/10.22331/q-2020-11-11-361}{{Quantum} {\bf
  4}, 361}~(2020).

\bibitem{CKS}
Andrew~M. Childs, Robin Kothari, and Rolando~D. Somma.
\newblock ``Quantum algorithm for systems of linear equations with
  exponentially improved dependence on precision''.
\newblock \href{https://dx.doi.org/10.1137/16M1087072}{SIAM Journal on
  Computing {\bf 46}, 1920--1950}~(2017).

\bibitem{CostaAnYuvalEtAl2022}
Pedro~C.S. Costa, Dong An, Yuval~R. Sanders, Yuan Su, Ryan Babbush, and
  Dominic~W. Berry.
\newblock ``Optimal scaling quantum linear-systems solver via discrete
  adiabatic theorem''.
\newblock \href{https://dx.doi.org/10.1103/PRXQuantum.3.040303}{PRX Quantum
  {\bf 3}, 040303}~(2022).

\bibitem{RobinAram}
Aram~W. Harrow and Robin Kothari.
\newblock ``{\relax}''.
\newblock In preparation~(2025).

\bibitem{Lin2020}
Lin Lin and Yu~Tong.
\newblock ``Optimal polynomial based quantum eigenstate filtering with
  application to solving quantum linear systems''.
\newblock \href{https://dx.doi.org/10.22331/q-2020-11-11-361}{Quantum {\bf 4},
  361}~(2020).

\bibitem{PhysRevLett.122.060504}
Yi{\u{g}}it Suba{\c{s}}{\i}, Rolando~D Somma, and Davide Orsucci.
\newblock ``Quantum algorithms for systems of linear equations inspired by
  adiabatic quantum computing''.
\newblock \href{https://dx.doi.org/10.1103/PhysRevLett.122.060504}{Physical
  Review Letters {\bf 122}, 060504}~(2019).

\bibitem{jennings2023efficient}
David Jennings, Matteo Lostaglio, Sam Pallister, Andrew~T Sornborger, and
  Yiğit Subaşı.
\newblock ``Efficient quantum linear solver algorithm with detailed running
  costs''~(2023).
\newblock
  url:~\href{https://arxiv.org/abs/2305.11352}{arxiv.org/abs/2305.11352}.

\bibitem{jansen2007bounds}
Sabine Jansen, Mary-Beth Ruskai, and Ruedi Seiler.
\newblock ``Bounds for the adiabatic approximation with applications to quantum
  computation''.
\newblock \href{https://dx.doi.org/https://doi.org/10.1063/1.2798382}{Journal
  of Mathematical Physics {\bf 48}, 102111}~(2007).

\bibitem{Sanders_2020}
Yuval~R. Sanders, Dominic~W. Berry, Pedro~C.S. Costa, Louis~W. Tessler, Nathan
  Wiebe, Craig Gidney, Hartmut Neven, and Ryan Babbush.
\newblock ``Compilation of fault-tolerant quantum heuristics for combinatorial
  optimization''.
\newblock \href{https://dx.doi.org/10.1103/prxquantum.1.020312}{{PRX} Quantum
  {\bf 1}, 020312}~(2020).

\bibitem{BabbushBerryNeven2019}
Ryan Babbush, Dominic~W. Berry, and Hartmut Neven.
\newblock ``Quantum simulation of the sachdev-ye-kitaev model by asymmetric
  qubitization''.
\newblock \href{https://dx.doi.org/10.1103/PhysRevA.99.040301}{Physical Review
  A {\bf 99}, 040301}~(2019).

\bibitem{berry2024doubling}
Dominic~W. Berry, Danial Motlagh, Giacomo Pantaleoni, and Nathan Wiebe.
\newblock ``Doubling the efficiency of hamiltonian simulation via generalized
  quantum signal processing''.
\newblock \href{https://dx.doi.org/10.1103/PhysRevA.110.012612}{Phys. Rev. A
  {\bf 110}, 012612}~(2024).

\bibitem{costa2024qlsp2}
Pedro C.~S. Costa.
\newblock ``Qlsp via discrete adiabatic method - source code''.
\newblock
  \url{https://github.com/PcostaQuantum/QLSP-via-discrete-adiabatic-method/blob/main/Walk_error_Herm.m}~(2025).
\newblock Accessed: 2025-01-24.

\bibitem{sparse_matrix_collection}
Tim Davis and Yifan Hu.
\newblock ``Suitesparse matrix collection''.
\newblock \url{https://sparse.tamu.edu/}~(2024).
\newblock Accessed: 2024-10-23.

\bibitem{costa2024qlsp}
Pedro C.~S. Costa.
\newblock ``Qlsp via randomisation method - source code''.
\newblock
  \url{https://github.com/PcostaQuantum/QLSP-via-randomisation-method}~(2024).
\newblock Accessed: 2025-01-24.

\bibitem{parity1}
Edward Farhi, Jeffrey Goldstone, Sam Gutmann, and Michael Sipser.
\newblock ``Limit on the speed of quantum computation in determining parity''.
\newblock \href{https://dx.doi.org/10.1103/PhysRevLett.81.5442}{Phys. Rev.
  Lett. {\bf 81}, 5442--5444}~(1998).

\bibitem{parity2}
Robert Beals, Harry Buhrman, Richard Cleve, Michele Mosca, and Ronald de~Wolf.
\newblock ``Quantum lower bounds by polynomials''.
\newblock \href{https://dx.doi.org/10.1145/502090.502097}{J. ACM {\bf 48},
  778–797}~(2001).

\end{thebibliography}
\end{document}